\documentclass[%
reprint,
superscriptaddress,
preprintnumbers,
nofootinbib,
amsmath,amssymb,
aps,
prd,
]{revtex4-1}

\usepackage{hyperref}
\hypersetup{colorlinks=true, citecolor=blue, urlcolor=blue, linkcolor=blue}
\usepackage[T1]{fontenc}
\usepackage{float}      		% figure positioning
\usepackage{dcolumn}    		% Align table columns on decimal point
\usepackage{bm}         		% bold math
\usepackage{graphicx}		% Including figure files
\usepackage{amsmath}		% Advanced maths commands
\usepackage{amssymb}		% Extra maths symbols
\usepackage{aas_macros}
\usepackage{xcolor}
\usepackage{newtxtext,newtxmath}

\begin{document}
\title{Growth and Geometry Split in Light of the DES-Y3 Survey}

\author{Kunhao Zhong}\email{kunhao.zhong@stonybrook.edu}
\affiliation{Department of Physics \& Astronomy, Stony Brook University, Stony Brook, NY 11794, USA}
\author{Evan Saraivanov}
\affiliation{Department of Physics \& Astronomy, Stony Brook University, Stony Brook, NY 11794, USA}
\author{Vivian Miranda}
\affiliation{Department of Physics \& Astronomy, Stony Brook University, Stony Brook, NY 11794, USA}
\affiliation{C. N. Yang Institute for Theoretical Physics, Stony Brook University, Stony Brook, NY, 11794, USA}
\author{Jiachuan Xu}
\affiliation{Department of Astronomy/Steward Observatory, University of Arizona, 933 North Cherry Avenue, Tucson, AZ 85721, USA}
\author{Tim Eifler}
\affiliation{Department of Astronomy/Steward Observatory, University of Arizona, 933 North Cherry Avenue, Tucson, AZ 85721, USA}
\author{Elisabeth Krause}
\affiliation{Department of Astronomy/Steward Observatory, University of Arizona, 933 North Cherry Avenue, Tucson, AZ 85721, USA}
\affiliation{Department of Physics, University of Arizona,  1118 E Fourth Str, Tucson, AZ, 85721-0065, USA}

\date{\today}

\begin{abstract}
We test the smooth dark energy paradigm using Dark Energy Survey (DES) Year 1 and Year 3 weak lensing and galaxy clustering data. Within the $\Lambda$CDM and $w$CDM model we separate the expansion and structure growth history by splitting $\Omega_\mathrm{m}$ (and $w$) into two meta-parameters that allow for different evolution of growth and geometry in the Universe. We consider three different combinations of priors on geometry from CMB, SNIa, BAO, BBN that differ in constraining power but have been designed such that the growth information comes solely from the DES weak lensing and galaxy clustering. 
For the DES-Y1 data we find no detectable tension between  growth and geometry meta-parameters in both the $\Lambda$CDM and $w$CDM parameter space. This statement also holds for DES-Y3 cosmic shear and 3x2pt analyses. For the combination of DES-Y3 galaxy-galaxy lensing and galaxy clustering (2x2pt) we measure a tension between our growth and geometry meta-parameters of 2.6$\sigma$ in the $\Lambda$CDM and 4.48$\sigma$ in the $w$CDM model space, respectively. We attribute this tension to residual systematics in the DES-Y3 \textsc{RedMagic} galaxy sample rather than to new physics. We plan to investigate our findings further using alternative lens samples in DES-Y3 and future weak lensing and galaxy clustering datasets. 
\end{abstract}

\maketitle

\section{Introduction}

Since the discovery of the accelerated expansion of our Universe~\cite{SupernovaSearchTeam:1998fmf, SupernovaCosmologyProject:1998vns}, the flat $\Lambda$CDM, which adopts a late-time Universe dominated by the cosmological constant, has become the standard model of cosmology. 
From a fundamental physics viewpoint, the origin of dark energy is still unknown. The cosmological constant modeled as vacuum energy is fine-tuned with a value too small to any known quantum field theory~\cite{Weinberg:1988cp}. Dynamical scalar fields, \textit{quintessence} and \textit{k-essence}, have been proposed to solve the fine-tuning problem~\cite{Caldwell:1997ii,Zlatev1999,Tsujikawa2013,Armendariz-Picon:2000ulo,Cai2010}. Modified gravity is an alternative way to explain the Universe's acceleration without introducing a new component~\cite{Tsujikawa2010MG}. To date, none of these proposed scenarios have been detected by observations. 

With only six free parameters, the standard model of cosmology predicts the temperature and polarization anisotropy statistics of the Cosmic Microwave Background (CMB) with remarkable success. Additionally, imaging and spectroscopic surveys show increasing power to constrain $\Lambda$CDM's predictions for the late-time evolution of large-scale structures (LSS); current stage III LSS surveys include the Dark Energy Survey (DES)~\cite{DES:2016jjg,DES:2017myr,DES:2017qwj,DES:2018nnc,DES:2020ahh,DES:2021wwk,DES:2021rex,DES:2022ygi,DES:2021bvc}, the Kilo-Degree Survey (KiDS)~\cite{Kuijken:2015vca, Joudaki:2019pmv,Heymans:2020gsg,KiDS:2020suj,Ruiz-Zapatero:2021rzl}, the Hyper Suprime-Cam Subaru Strategic Program (HSC)~\cite{aihara2018hyper,HSC:2018mrq,Hamana:2019etx,2022arXiv221116516R,2022PASJ...74..247A}, and the Baryon Oscillation Spectroscopic Survey (Boss and eBOSS)~\cite{Smee2013,Dawson:2015wdb,eBOSS:2020yzd,2021MNRAS.506.2503M,2022MNRAS.511.5492Z,2022MNRAS.516..617C}.

However, multiple tensions have arisen in the last few years within the $\Lambda$CDM model, particularly between Planck measurements of the Cosmic Microwave Background and data from the late-time Universe. The first tension involves the value of the Hubble constant, $H_0$~\cite{Douspis:2018xlj,Riess:2019cxk,Wong:2019kwg,sh0es21}. Local-Universe $H_0$ estimates from type Ia supernova (SNIa), calibrated using Cepheid variable stars~\cite{Riess:2016jrr,Riess:2020sih}, conflict with CMB predictions~\cite{Ade:2015xua,Planck:2018vyg}. Several studies show that this tension is reaching a statistical significance of $5\sigma$~\cite{sh0es21,Riess:2019cxk,Wong:2019kwg}.

Hubble constant predictions from the Cosmic Microwave Background are sensitive to changes in the late-time dark sector~\cite{Hu:2004kn}. For example, cold dark matter models decaying to relativistic species can affect the CMB predictions~\cite{Ade:2015rim,Pandey:2019plg,Clark:2020miy}. These predictions are also sensitive to physics before recombination via the sound horizon. However, observations of SNIa combined with Baryonic Acoustic Oscillations (BAO) show that changes in the late-time Universe dark sector cannot solve the $H_0$ tension without creating additional problems~\cite{Addison:2017fdm,Lemos:2018smw,Dhawan:2020xmp}. These constraints suggest that the new physics should come from the time before recombination~\cite{Verde:2019ivm,Knox:2019rjx}.

The Dark Energy Survey year one (DES-Y1) and year three (DES-Y3) analysis conclude that the parameter $S_8$ is in mild tension with the $\Lambda$CDM model predicted by Planck CMB data~\cite{DES:2017hdw,DES:2021zxv,Y3MagLim}. Multiple independent surveys have independently discovered this discrepancy~\cite{Mantz:2014paa,Hildebrandt:2016iqg,Joudaki:2019pmv,eBOSS:2020yzd}. The projected one-dimensional $S_8$ tension is not large; however, investigations of the multi-dimensional degeneracy directions in $\Lambda$CDM parameter space offers a more complete picture~\cite{Secco:2022kqg}. The generalizations of the late-time dark sector can reduce this discrepancy, but the $S_8$ tension generally increases with statistical significance when an early-dark energy component is added in the $\Lambda$CDM model~\cite{Hill:2020osr,Ivanov2020,2021JCAP...05..072D}.

In this work, we split the matter density, $\Omega_\mathrm{m}$, and the dark energy equation of state, $w$, to test the consistency of smooth-dark-energy between the background evolution and the late-time scale-independent growth of structures~\cite{Wang:2007fsa,Mortonson:2008qy,Mortonson_2010,2015PhRvD..91f3009R,Miranda:2017mnw}. Using different data sets containing geometry or growth information, we can verify such consistency in $\Lambda$CDM and $w$CDM models. Parameter splitting has been extensively applied in multiple contexts. For example, baryon density can be divided into two parts with one only affecting ionization history~\cite{Chu:2004qx}, cold matter density can be split into parts representing different aspects of type Ia supernova~\cite{10.1111/j.1745-3933.2008.00519.x}, or the primordial inflationary amplitude can be separated into one that affects the CMB and another that only affects predictions from the effective field theory of large-scale structure~\cite{Smith:2020rxx}.

This work is a follow-up investigation of two previous analyses, one employing DES-Y1 data~\cite{DES:2020iqt}, and the other adopting older weak lensing data from the Canada-France Hawaii Telescope Lensing Survey~\cite{2015PhRvD..91f3009R}. In this work, we employ the new  DES-Y3 3x2pt data, including different data combinations that clarify some internal aspects of the galaxy-galaxy lensing and galaxy clustering combination. The Kilo-Degree Survey (KiDS) collaboration also analyzed their data with the growth-geometry split type of parameters~\cite{Ruiz-Zapatero:2021rzl}. In addition to weak lensing and galaxy clustering, redshift space distortion (RSD) and clusters data are used to extract growth information~\cite{Bernal:2015zom,Andrade:2021njl}. Previous weak lensing work with DES-Y1 data~\cite{DES:2020iqt} does not report a disagreement with the $\Lambda$CDM model. However, RSD data do favor a lower growth rate. See Sec.~\ref{sec:Conclusion} for a detailed discussion.

The structure of the paper is as follows: In Sec.~\ref{sec:methodology}, we explain the geometry-growth split and the 3x2pt combination of two-point correlation functions. We summarize DES analysis choices and the external data sets in Sec.~\ref{sec:data_and_analysis_method}, which also contains a detailed description of our adopted pipeline and the validation tests we performed based on synthetic $\Lambda$CDM DES-Y1 and DES-Y3 data vectors. We present the results and discussions in Sec.~\ref{sec:results}, and conclusions, including an exposition on planned follow-up improvements, in Sec.~\ref{sec:Conclusion}.

\section{Theory and Methodology}
\label{sec:methodology}

\subsection{Split Matter Power Spectrum}
\label{section: Split Matter Power Spectrum}

The linear matter power spectrum quantifies the inhomogeneity of matter distribution, and it can be written as the product of the inflationary primordial spectrum, the transfer function, and the growth function:
\begin{multline}\label{def:Pk}
P^{\rm linear}(z, k)=\frac{2 \pi^2}{k^3} \frac{4}{25} A_{\mathrm{s}}\left(\frac{k}{k_{\mathrm{norm}}}\right)^{n_{\mathrm{s}}-1}\left(\frac{k}{H_0}\right)^4 \, \times \\ \times \,  T^2(k) \left(\frac{G(z)}{\Omega_{\mathrm{m}}(1+z)}\right)^2
\end{multline}
The growth function, 
\begin{equation}
G(z) = (1+z) D(z) = (1+z) \dfrac{\delta_\mathrm{m}(z)}{\delta_\mathrm{m}(z_{\rm ini})},
\end{equation}
describes the scale-independent time evolution of matter overdensity from initial conditions defined at redshift $z_{\rm ini}=1000$. In smooth dark energy cosmologies, the growth-factor evolution obeys the following ordinary differential equation:
\begin{align}\label{eq:g-growth}
G^{\prime \prime}+ \left(4+\frac{H^{\prime}}{H}\right) G^{\prime}+\left[3+\frac{H^{\prime}}{H}-\frac{3}{2} \Omega_{\mathrm{m}}(z)\right] G=0 , 
\end{align}
where the prime denotes derivative with respect to the logarithm of the scale factor, $\ln a$. The initial conditions are $G_{\rm ini} = 1$ and $G'_{\rm ini} = -(3/5) (1-w) \Omega_{\rm DE}(z_{\rm ini})$~\cite{Mortonson:2008qy}. Models that introduce clustering of dark energy break this scale-independent relation between growth factor and dark energy parameters~\cite{Ishak:2005zs, Huterer:2006mva}. In this work, we confine our study to the case of smooth dark energy with a constant equation of state ($w$CDM). Our results can be generalized, for example, by considering instead principal component based $w(z)$ parameterizations~\cite{Mortonson2010PhRvD..82f3004M, Miranda:2017mnw}. 

We split the $\Omega_\mathrm{m}$ and $w$ parameters into geometry, $\{\Omega_\mathrm{m}^{\rm geo}, w^{\rm geo}\}$, and growth counterparts $\{\Omega_\mathrm{m}^{\rm growth}, w^{\rm growth}\}$. The growth parameters affect the late-time growth factor evolution via Eq.~\ref{eq:g-growth}. The remaining parameters, $\{ \Omega_\mathrm{b}, H_0, A_\mathrm{s}, n_\mathrm{s}, \tau \}$, are not split. The split $\Lambda$CDM cosmology assumes $w^{\rm geo} = w^{\rm growth} = -1$. Since the linear power spectrum $P^{\rm linear}(z, k) \propto G^2(z)$, we can define the split linear matter power spectrum to be:

\begin{align}\label{def:Pk-split-new}
P_{\rm split}^{\rm linear}(k , z) =  \dfrac{P_{\rm geo}^{\rm linear-camb}(k,z)}{G_{\rm geo}^{\rm camb}(z)^2} \times G_{\rm growth}(z)^2 \, ,
\end{align}
with
\begin{align}\label{def:G_camb}
G_{\rm growth}(z) = G_{\rm geo}^{\rm camb}(z) \times \bigg(\dfrac{G_{\rm growth}^{\rm ODE}(z)}{G_{\rm geo}^{\rm ODE}(z)}\bigg) \, .
\end{align}  
$P_{\rm geo}^{\rm linear-camb}$ and $G_{\rm geo}^{\rm camb}$ are respectively the linear power spectrum and the growth factor both computed by the Boltzmann code CAMB~\cite{Lewis:1999bs,Howlett:2012mh} assuming the geometry parameters. $G_{\rm geo}^{\rm ODE}$ and $G_{\rm growth}^{\rm ODE}$ are solutions of the differential Eq.~\ref{eq:g-growth} given geometry and growth parameters respectively.

Our slightly convoluted definition is analytically equivalent to
\begin{align}\label{def:Pk-slpit}
P_{\rm split}^{\rm linear}(k , z) =  P_{\rm geo}^{\rm linear-camb}(k,z) \times  \bigg(\dfrac{G_{\rm growth}^{\rm ODE}(z)}{G_{\rm geo}^{\rm ODE}(z)}\bigg)^2  \, .
\end{align}
Definition of Eq.~\ref{def:G_camb} resolves the small numerical error between the growth factor calculated by CAMB versus the solution from Eq.~\ref{eq:g-growth} with no radiation and accurate background evolution of massive neutrinos; the adopted multi-probe lensing pipeline requires $G_{\rm growth}(z)$ itself when computing intrinsic alignment contributions for cosmic shear and galaxy-galaxy lensing.
\begin{figure}[t]
\includegraphics[width=\columnwidth]{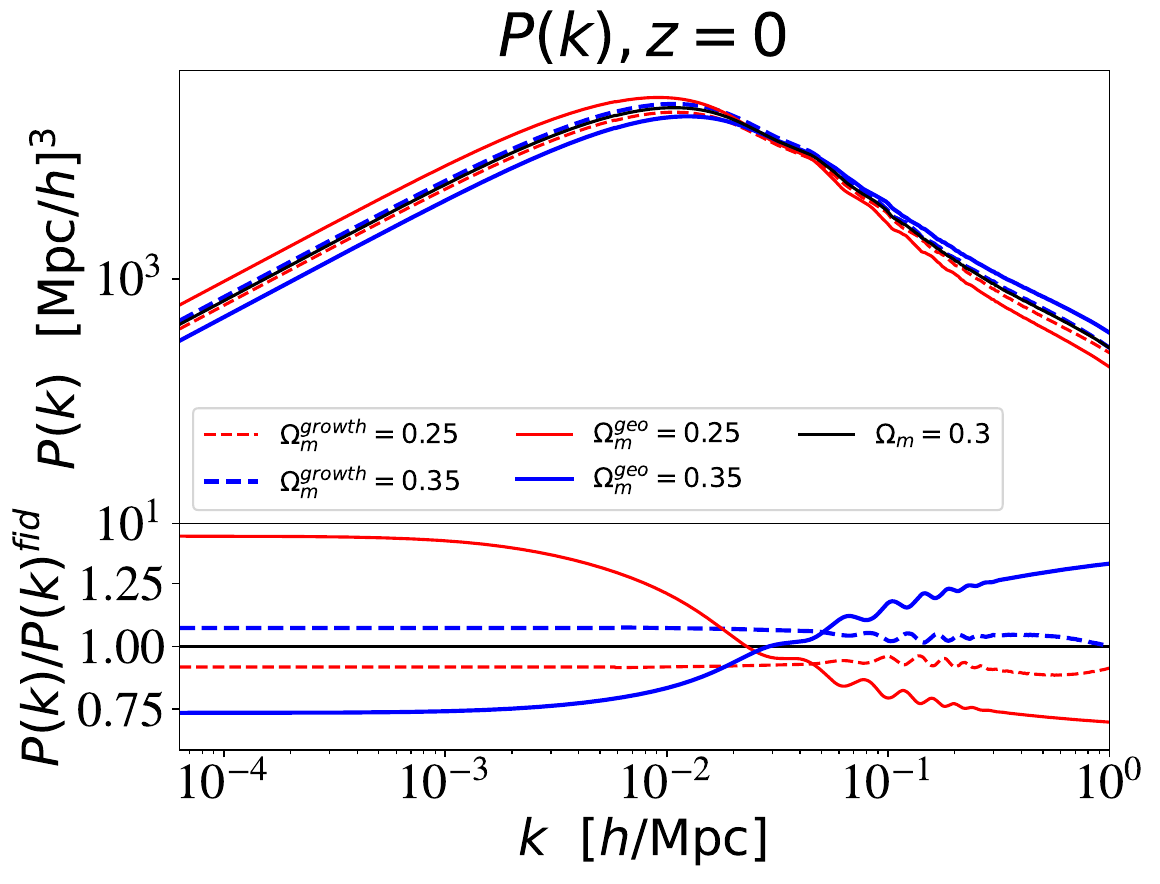}
\caption{Geometry and growth effects on the matter power spectrum in the split-$\Lambda$CDM model. When changing one parameter, we keep the remaining matter density at $\Omega_{\rm m}^{\rm X}$ = 0.3. The choice of employing $\Omega_{\rm m}^\mathrm{\rm growth}$ in the Euclid Emulator has the effect of roughly maintaining the scale-independent amplitude shift induced by changes in $\Omega_{\rm m}^\mathrm{\rm growth}$ on mildly non-linear scales. On $k \gtrsim 10^{-2}$ scales that are within DES reach, changes in the shape parameter $\Gamma = \Omega_{\rm m}^{\mathrm{geo}} h$ induced by varying $\Omega_{\rm m}^{\mathrm{geo}}$ are degenerate with the primordial power spectrum tilt, $n_{\mathrm{s}}$. This degeneracy motivates our choice of priors; CMB brings external constraints on the inflationary and shape parameters, while BAO and SNIa indirectly limits the shape parameter.}
\label{fig:pksplit}
\end{figure}

We follow the naming convention in the parameter split literature. In our parameter split distance probes heavily constrain geometry parameters while growth parameters allow the late-time growth factor to vary with extra degrees of freedom. However $\Omega_\mathrm{m}^{\rm geo}$ and $w^{\rm geo}$ can also affect structure growth. Specifically, the split matter power spectrum in our definition is proportional to $(\Omega_\mathrm{m}^{\rm geo})^{-2}$ (see Fig.~\ref{fig:pksplit}). Additionally, early universe physics that affect both background expansion and structure formation are also modeled by $\Omega_\mathrm{m}^{\rm geo}$. The split between geometry and growth is not uniquely defined, and we defer to future work examining the impact of these choices. 
\begin{figure}[t]
\centering
\includegraphics[width=\columnwidth]{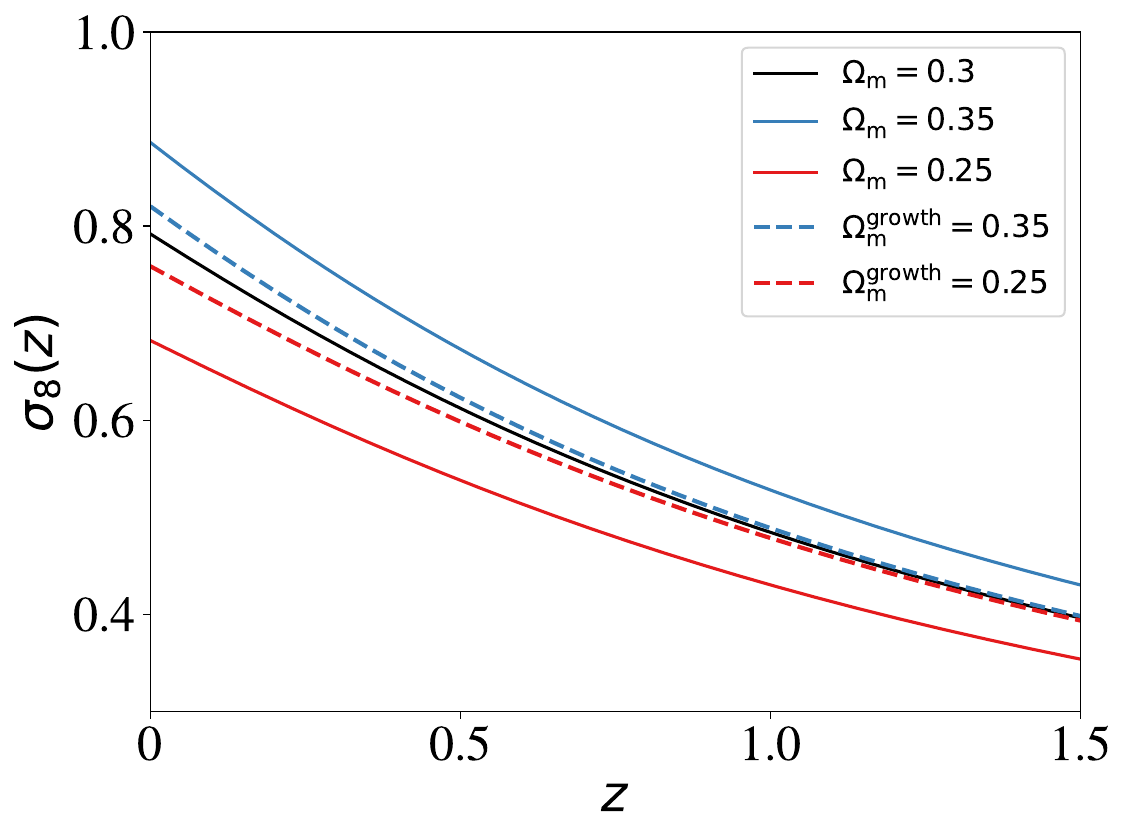}
\includegraphics[width=\columnwidth]{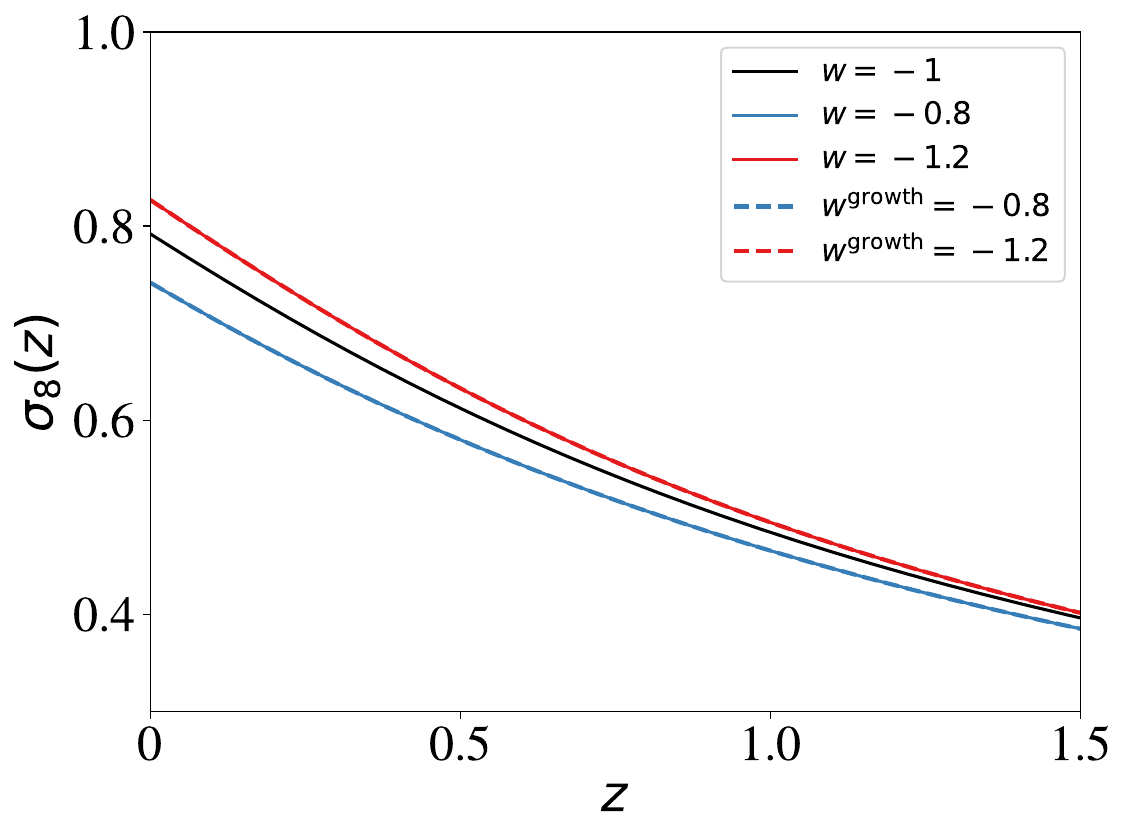}
\caption{Both panels show $\sigma_8$ changes under the variations of the unsplit and split $\Omega_{\rm m}$ and $w$ parameters. The solid lines show shifts on the unsplit $\Lambda$CDM and $w$CDM models, and the dashed lines change growth while keeping the geometry parameters at their fiducial $\Omega_{\rm m}^\mathrm{geo} = 0.3$ and $w^\mathrm{geo}$ = -1. We can see that changes in the growth parameter $\Omega_\mathrm{m}^{\rm growth}$ result in a minor shift in $\sigma_8(z)$, whereas $w^{\rm growth}$ gives the same change as in the non-split case because it only affects the matter power spectrum through the growth function.}
\label{fig:sigma8} 
\end{figure}
\\ \\
The root mean variance within 8 Mpc/h is defined as:
\begin{align}
\sigma_{8}^{2}(z)=\frac{1}{2 \pi^{2}} \int \mathrm{d} \log k W^{2}(k R) k^{3} P(k, z),
\end{align}
where $W(kR)$ is a top-hat filter function in Fourier space with radius $R=8$ Mpc/h. The split $\sigma_8^{\rm split}(z)$ is then given by:
\begin{align}
\sigma_8^{\rm split}(z) = \sigma_8^{\rm geo-camb}(z) \times \bigg(\dfrac{G_{\rm growth}^{\rm ODE}(z)}{G_{\rm geo}^{\rm ODE}(z)}\bigg).
\end{align}
The different behavior of $\sigma_8^{\rm split}(z)$ versus $\sigma_8(z)$ with respect to the change of growth and geometry parameters is shown in Fig.~\ref{fig:sigma8}. In the split $\Lambda$CDM case, the change of $\Omega_\mathrm{m}^{\rm growth}$ will give a smaller change on $\sigma_8(z)$ compared with the non-split case, namely when changing $\Omega_\mathrm{m}^{\rm growth}$ and $\Omega_\mathrm{m}^{\rm geo}$ simultaneously. In the split $w$CDM case a change in $w^{\rm growth}$ gives the same change as in the non-split case. In both cases, the change is larger at low redshift. 

\begin{figure}[t]
\includegraphics[width=\columnwidth]{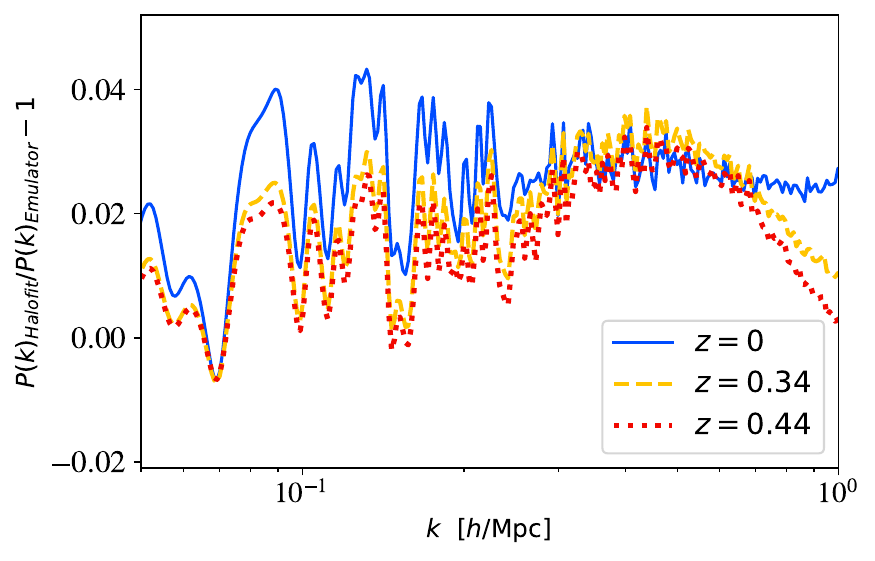}
\includegraphics[width=\columnwidth]{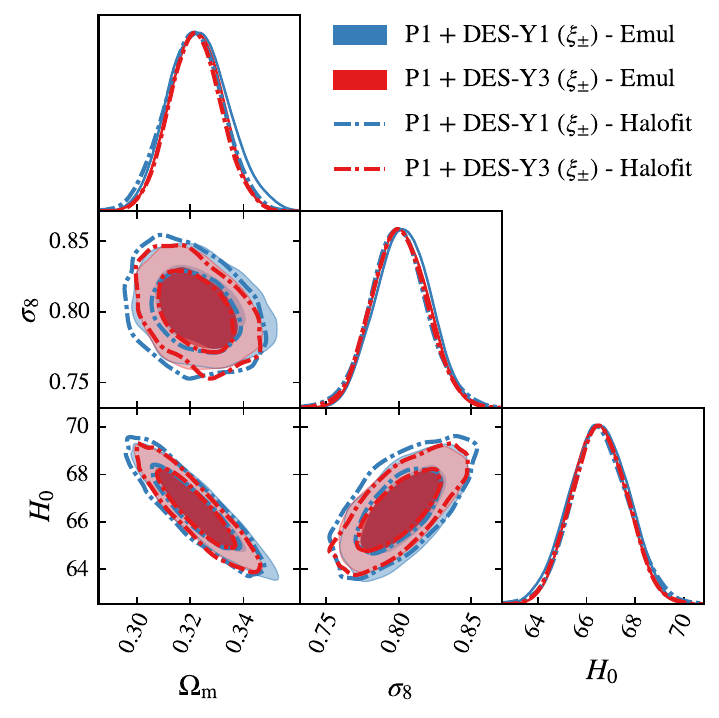}
\caption{\textit{Top panel}: Fractional difference of non-linear power spectrum between \textsc{Halofit} and the Euclid Emulator at three different redshifts; Fiducial parameter values are the same adopted in the synthetic DES chains (see Sec.~\ref{sec:validation}). \textit{Bottom panel}: posterior comparison between  \textsc{Halofit} and the Euclid Emulator on P1 + DES-Y1/Y3 cosmic shear combinations (P1 prior is defined in Sec.~\ref{sec:data_set_for_geometry}). We assumed the $\Lambda$CDM model in all four chains, and the DES-Y1/Y3 data vector was synthetic with the same fiducial model adopted in Sec.~\ref{sec:validation}. Both figures illustrate that within the prior ranges adopted in this manuscript, the few percent differences between \textsc{Halofit} and the \textsc{Euclid Emulator} do not impact our results.
}
\label{fig:halo_vs_emu}
\end{figure}

To account for non-linearities in the split power spectrum, we utilize the \textsc{Euclid Emulator} to compute the factor $B(k,z) \equiv P(k,z)\big/P^{\rm linear}(k,z)$~\cite{Euclid:2020rfv}. In this work, we defined the $B(k,z)$ as dependent on the growth parameters. We then define the split matter power spectrum as 
\begin{align}\label{eq:pk_spit2}
P_{\rm split}(k, z) =  P_{\rm split}^{\rm linear}(k, z) \times B^{\rm growth}(k, z) \, .
\end{align}
    The official DES-Y1 and DES-Y3 analyses adopt \textsc{Halofit}. Figure~\ref{fig:halo_vs_emu} shows that \textsc{Halofit} and \textsc{Euclid Emulator} differences are within 5\%. This disagreement doesn't affect inferences on the $\Lambda$CDM parameters as shown in the bottom panel of Fig.~\ref{fig:halo_vs_emu}. However, our definition has practical advantages. 

Massive neutrinos break the scale-independent evolution of dark matter perturbations; neutrinos transition from relativistic to non-relativistic behavior as the universe cools down. The scale-dependant changes in the matter spectrum are absorbed in $P_{\rm geo}^{\rm linear}(k,z)$ calculated by the Boltzmann code. For this paper, we only study fixed neutrino mass with $\sum_{\nu} m_{\nu} = 0.06\mskip\thinmuskip \mathrm{eV}$.

\section{Two-Point Correlation Functions}
\label{sec:Weak_Lensing_and_Galaxy_clustering}

\subsection{Weak Lensing and Galaxy clustering }
\label{sec:two-point_correlation_functions}

The dark matter distribution of the universe is traced by two fields: i) the galaxy density field, and ii) the weak lensing shear field. These fields generate three two-point correlation functions (2PCF) as a function of angular separation $\theta$: \\ \\
\textbf{Cosmic shear:} $\xi^{ij}_{\pm}(\theta)$: the correlation, $\langle \kappa \kappa \rangle$, between source galaxy shear in redshift bins $i$ and $j$. \\ \\
\textbf{Galaxy-Galaxy lensing} $\gamma^{ij}(\theta)$: the correlation, $\langle \delta_g \kappa \rangle$,  between lens galaxy positions and source galaxy tangential shear in redshift bins $i$ and $j$. \\ \\
\textbf{Galaxy Clustering} $ w^{ij}(\theta) $: the correlation,  $\langle \delta_g \delta_g \rangle$, between lens galaxy position in redshift bins $i$ and $j$. \\ \\
In combination, these probes significantly increase the information about the matter distribution and improve the systematics mitigation. Throughout this paper, "3x2pt" refers to the multi-probe analysis involving the combination of the three 2PCF, and "2x2pt" refers to the multi-probe combination of galaxy-galaxy lensing and galaxy clustering ($\gamma_t + w_\theta$). 

Theory predictions and 2PCF are related by the angular power spectra. In both DES-Y1 and DES-Y3, we calculate the full non-Limber integral on large angles only in the galaxy position auto power spectra, following~\citet{Fang:2019xat}. Using Limber approximation, the angular power spectra of tracer $A$ at redshift bin $i$ and tracer $B$ at redshift bin $j$ is~\cite{1953ApJ...117..134L, LoVerde:2008re}:
\begin{equation}\label{eq: C_AB limber approx}
C_{A B}^{i j}(\ell)=\int d \chi \frac{W_{A}^{i}(\chi) W_{B}^{j}(\chi)}{\chi^{2}(z)} P_{\rm split} \left( k , z (\chi) \right) \Big|_{k = (l+1/2)/\chi},
\end{equation}
where $\chi$ is the comoving radial distance. The weighting function of weak lensing shear $\kappa$ and galaxy number density $\delta_g$ are respectively~\cite{Bartelmann:1999yn}
\begin{equation}\label{eq: Window of shear}
W_{\kappa}^{i}(\chi)=\frac{3 H_{0}^{2} \Omega_{\mathrm{m}}^{\rm geo}}{2 \mathrm{c}^{2}} \frac{\chi}{a(\chi)} \int_\chi^{\infty} \mathrm{d} \chi^{\prime} \frac{n_{\kappa}^{i}\left(z\left(\chi^{\prime}\right)\right) d z / d \chi^{\prime}}{\bar{n}_{\kappa}^{i}} \frac{\chi^{\prime}-\chi}{\chi^{\prime}}
\end{equation}
and
\begin{equation}\label{eq: Window of galaxy}
W_{\delta_{\mathrm{g}}}^{i}(\chi)=b^{i}(z(\chi)) \frac{n_{\mathrm{g}}^{i}(z(\chi))}{\bar{n}_{\mathrm{g}}^{i}} \frac{d z}{d \chi} \, .
\end{equation}
Here, $\bar{n}_{\mathrm{g} / \kappa}^{i}=\int dz \, n_{\mathrm{g} / \kappa}^{i}(z)$ is the angular number density of galaxies in the redshift bin $i$, and $b^{i}(z(\chi))$ is the galaxy bias. Geometry parameters model the comoving radial distance~\cite{DES:2020iqt}. Being consistent with Eq.~\ref{def:Pk}, $P(k) \propto (1 \big/ \Omega_\mathrm{m}^{\rm geo})^2$, the matter density that appears in the $W_{\kappa}$ prefactor is $\Omega_\mathrm{m}^{\rm geo}$. This choice mainly follows the preferences of ~\cite{DES:2020iqt}. We defer the interesting investigation of how changing $\Omega_\mathrm{m}^{\rm geo} \to \Omega_\mathrm{m}^{\rm growth}$ here would affect the comparison between growth and geometry parameters.

The relation between two-point correlation functions and angular power spectra assumes bin-average curved sky formulas in both DES-Y1 and DES-Y3 as shown below
\begin{align}\label{eq: relation of 2PCF and angular power spectra}
w_{\theta}^{i}(\overline{\theta}) =& \sum_\ell \frac{2\ell+1}{4\pi}\overline{P_\ell} C^{ii}_{\delta\delta}(\ell)\, ,\\
\gamma_t^{ij}(\overline{\theta}) =& \sum_\ell \frac{2\ell+1}{4\pi\ell(\ell+1)}\overline{P^2_\ell} C^{ij}_{\delta\mathrm{E}}(\ell)\, , \\\nonumber
\xi_{\pm}^{ij}(\overline{\theta}) = & \sum_\ell\frac{2\ell+1}{2\pi\ell^2(\ell+1)^2}\overline{[G_{\ell,2}^+\pm G_{\ell,2}^-]}\left[
 C^{ij}_{EE}(\ell)\pm C^{ij}_{BB}(\ell)\right]\, ,
\end{align}
The analytical expressions for Legendre and associated Legendre polynomials $\overline{P_\ell}$ and $\overline{G_{\ell,2}^\pm}$ can be found in ~\cite{1996astro.ph..9149S}. Further information about these transformations, including $E/B$-mode projections on the auto and cross power spectra involving shear, are described in the DES-Y3 methods paper~\cite{DES:2021rex}. The computation of non-limber integrals in galaxy position auto power spectra, and the use of bin-average curved sky formulas for cosmic shear, galaxy-galaxy lensing and galaxy clustering in DES-Y1 is an improvement over modeling choices of  ~\cite{DES:2020iqt}. 

We use the Tidal Alignment and Tidal Torquing (TATT), a generalization to the previously DES-Y1 adopted non-linear alignment model (NLA-IA), to model the intrinsic alignment of galaxies in DES-Y3 data~\cite{Blazek:2017wbz,DES:2017tss,DES:2021rex}. Under this framework, the intrinsic shape of galaxies is written as a collection of terms depending on the matter overdensity, $\delta_m$, and the tidal tensor, $s_{i j}$. These terms describe tidal alignment, tidal torquing, and density weighting, as shown below:
\begin{align}\label{eq: TATT}
\gamma_{i j}^{I}=\underbrace{C_{1} s_{i j}}_{\text{\tiny Tidal Alignment }}+\underbrace{b_\mathrm{TA} C_{1}\left(\delta_m \times s_{i j}\right)}_{\text {\tiny Density Weighting }}+\underbrace{C_{2}\left[s_{i }^{\, \, k} s_{k j}-\frac{1}{3} \delta_{i j} s^{2}\right]}_{\text {\tiny Tidal Torquing }} \, .
\end{align}
Here, 
\begin{align}
C_{1} &= -\frac{A_{1} \bar{C} \Omega_{\rm m}^{\mathrm{growth}}}{aG^{\mathrm{growth}}(z)}\left(\frac{1+z}{1+z_0}\right)^{\eta_{1}} \, , 
\end{align}
and 
\begin{align}
C_{2} &= 5\frac{A_{2} \bar{C}\Omega_{\rm m}^{\mathrm{growth}}}{(a G^{\mathrm{growth}}(z))^2}\left(\frac{1+z}{1+z_0}\right)^{\eta_{2}}\, .
\end{align}
The redshift $z_0$ is to the mean redshift of the source galaxy sample, and $\bar{C}= (5\times10^{-14} h^{-2}M_\odot^{-1} \mathrm{Mpc}^3) \times \rho_\mathrm{crit}$. The TATT model contains five parameters: the amplitude $A_{i=1,2}$, power law index $\eta_{i=1,2}$, and the effective source galaxy bias $b_{TA}$. The TATT model reduces to NLA-IA when $A_2 = b_{TA} = 0$. Both NLA-IA and TATT have an explicit dependence on matter density and the growth factor; we assume these are both growth parameters.

\begin{figure}[t]
\includegraphics[width=\columnwidth]{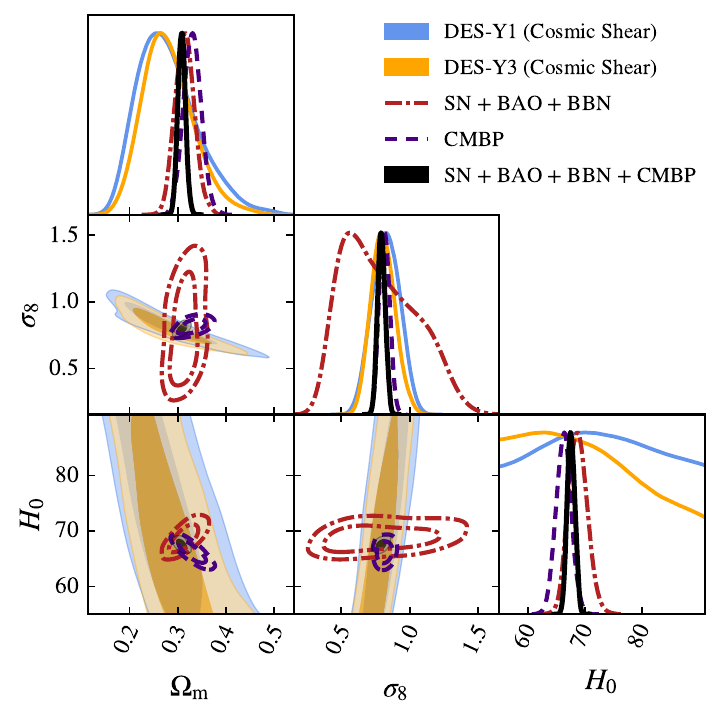}
\caption{Cosmic Shear posteriors in $\Lambda$CDM model for DES-Y1 and DES-Y3. Unlike all remaining figures and results in this manuscript, these constraints assume \textsc{Halofit} for the non-linear matter power spectrum and the original DES-Y3 priors for the cosmological parameters~\cite{DES:2021wwk}. The red dot-dashed lines are cosmological constraints from type Ia Supernova, BAO and BBN external data. On the other hand, the blue dashed lines show posteriors derived from the Cosmic Microwave Background temperature and polarization Planck 2018 data with the reduced multipole range $35<\ell<396$ in combination with low-$\ell$ EE polarization data $\ell$ < 30. These are the external data combinations adopted on priors P2 and P1, respectively (see Sec.~\ref{sec:data_set_for_geometry}). P1 and P2 priors are not stringent enough to create a significant $\sigma_8$ tension, but at the same time, they provide complementary information in parameters that DES does not constrain. While both priors measure $\Omega_{\rm m}$ and the Hubble constant $H_0$, only the CMB data restricts the inflationary parameters $A_{\mathrm{s}}$ and $n_{\mathrm{s}}$. As shown in ref.~\cite{DES:2020iqt}, external (non-DES) information is necessary to tightly constrain growth parameters. In comparison with the full TT+low-$\ell$ EE, the truncated CMB primary here has 3-4 larger standard deviation in $A_{\mathrm{s}}$ and $n_{\mathrm{s}}$, but Prior 1+2 is almost same constraining as full CMB in $\Omega_{\rm m}$}
\label{fig:prior test1}
\end{figure}

\section{Data and Analysis Method}
\label{sec:data_and_analysis_method}
\begin{table}[h]
\renewcommand{\arraystretch}{1.3}
\centering
\begin{tabular}{lr} % 
\hline
Cosmological Parameters & Prior \\
\hline\hline
$\Omega_\mathrm{m}^{\rm geo}$ &  Flat(0.1, 0.9) \\
$w^{\text{\rm geo}}$ & Flat(-3, -0.01) \\
$ A_{\mathrm{s}} \times 10^{9} $ & Flat(1.7, 2.5) \\
$ n_\mathrm{s} $ & Flat(0.92, 1.0) \\
$ H_0 $ & Flat(61, 73) \\
$ \tau $ & Flat(0.01, 0.8) \\
$\Omega_\mathrm{m}^{\text{\rm growth}}$ & Flat(0.24 , 0.4) \\
$ w^{\text{\rm growth}}  $ & Flat(-1.7 , -0.7) \\
\hline
\end{tabular}
\caption{Flat priors for the cosmological parameters. We take the priors as in the \textsc{Euclid Emulator} for the parameters $(A_\mathrm{s}, n_\mathrm{s}, H_0, \Omega_\mathrm{m}^{\text{\rm growth}}, w^{\text{\rm growth}})$. We only include the optical depth of reionization, $\tau$, in chains with CMB data.}
\label{table:prior_choices_cosmology}
\end{table}

\begin{table}
\renewcommand{\arraystretch}{1.3}
\centering
\begin{tabular}{lr} % 
\hline
DES-Y3 Nuisance Parameters &  Prior \\
\hline\hline
\textbf{Linear Galaxy bias} \\
$ b_g^i(i \in [1,5])$ &  Flat(0.8, 3.0) \\
\hline
\textbf{Intrinsic Alignment (TATT)} \\
$A_{1}$ &  Flat(-5, 5) \\
$A_{2}$ &  Flat(-5, 5) \\
$\eta_{1}$ &  Flat(-5, 5) \\
$\eta_{2}$ &  Flat(-5, 5) \\
$b_{TA}$ &  Flat(0 , 2) \\
\hline
\textbf{Source photo-z} \\
$\Delta z_{\mathrm{s}}^{1} \times 10^{2}$ &   Gauss(0, 1.8) \\
$\Delta z_{\mathrm{s}}^{2} \times 10^{2}$ &   Gauss(0, 1.5) \\
$\Delta z_{\mathrm{s}}^{3} \times 10^{2}$ &   Gauss(0, 1.1)\\
$\Delta z_{\mathrm{s}}^{4} \times 10^{2}$ &   Gauss(0, 1.7)\\
\hline
\textbf{Lens photo-z}\\
$\Delta z_{\mathrm{1}}^{1} \times 10^{2}$ &   Gauss(0.6, 0.4)  \\
$\Delta z_{\mathrm{1}}^{2} \times 10^{2}$ &   Gauss(0.1, 0.3)  \\
$\Delta z_{\mathrm{1}}^{3} \times 10^{2}$ &   Gauss(0.4, 0.3)\\
$\Delta z_{\mathrm{1}}^{4} \times 10^{2}$ &   Gauss(-0.2, 0.5)\\
$\Delta z_{\mathrm{1}}^{5} \times 10^{2}$ &   Gauss(-0.7, 0.1)\\
\hline
\textbf{Multiplicative shear calibration} \\
$m_{1} \times 10^2$ &   Gauss(-0.6, 0.9)\\
$m_{2} \times 10^2$ &   Gauss(-2.0, 0.8)\\
$m_{3} \times 10^2$ &   Gauss(-2.4, 0.8)\\
$m_{4} \times 10^2$ &   Gauss(-3.7, 0.8)\\
\hline
\textbf{Lens magnification} \\
$C_{\mathrm{1}}^1 \times 10^2$ &   Fixed (0.63)\\
$C_{\mathrm{1}}^2 \times 10^2$ &   Fixed (-3.04)\\
$C_{\mathrm{1}}^3 \times 10^2$ &   Fixed (-1.33)\\
$C_{\mathrm{1}}^4 \times 10^2$ &   Fixed (2.50)\\
$C_{\mathrm{1}}^5 \times 10^2$ &   Fixed (1.93)\\
\hline
\textbf{Point mass marginalization} \\
$B_i(i \in [1,5])$  & Flat(-5, 5) \\
\hline
\end{tabular}
\caption{Adopted priors on DES-Y3 nuisance parameters. The priors are either Flat (min, max) or Gaussian (mean, standard deviation).}
\label{table:prior_choices_Y3}
\end{table}

\begin{table}[t]
\renewcommand{\arraystretch}{1.3}
\centering
\begin{tabular}{l r}  
\hline
DES-Y1 Nuisance Parameters &  Prior \\
\hline
\hline
\textbf{Linear Galaxy bias} \\
$ b_g^i(i \in [1,5])$ &  Flat(0.8, 3.0) \\
\hline
\textbf{Intrinsic Alignment (NLA)} \\
$A_{1}$ &  Flat(-5, 5) \\
$\eta_1$ &  Flat(-5, 5) \\
\hline
\textbf{Source photo-z} \\
$\Delta z_{\mathrm{s}}^{1} \times 10^{2}$ & Gauss(-0.1, 1.6) \\
$\Delta z_{\mathrm{s}}^{2} \times 10^{2}$ & Gauss(-0.19, 1.3) \\
$\Delta z_{\mathrm{s}}^{3} \times 10^{2}$ & Gauss(0.9, 1.1)\\
$\Delta z_{\mathrm{s}}^{4} \times 10^{2}$ & Gauss(-1.8, 2.2)\\
\hline
\textbf{Lens photo-z}\\
$\Delta z_{\mathrm{1}}^{1} \times 10^{2}$ & Gauss(0.8, 0.7) \\
$\Delta z_{\mathrm{1}}^{2} \times 10^{2}$ & Gauss(-0.5, 0.7)  \\
$\Delta z_{\mathrm{1}}^{3} \times 10^{2}$ & Gauss(0.6, 0.6)\\
$\Delta z_{\mathrm{1}}^{i} \times 10^{2} (i \in [4,5]) $ &   Gauss(0, 0.01)\\
\hline
\textbf{Multiplicative shear calibration} \\
$m_{i} \times 10^2 (i \in [1,4])$ &   Gauss(1.2, 2.3)\\
\hline
\end{tabular}
\caption{Adopted priors on DES-Y1 nuisance parameters. The priors are either Flat (min, max) or Gaussian (mean, standard deviation). Note that the parameters not presented here correspond to systematics not considered for DES-Y1 analysis.}
\label{table:prior_choices_Y1}
\end{table}

\subsection{DES data}\label{sec: Data Set for DES}

This work presents results using DES-Y1 and DES-Y3 data; regarding DES-Y1~\cite{DES:2020iqt}, we have implemented significant changes in the choice of external data sets and non-linear modeling. In both data sets, the collaboration measured 2PCF via the~\textsc{TreeCorr} algorithm \cite{Jarvis:2003wq}. We follow the collaboration choices when applying scale cuts to remove small-scale information. The resulting 3x2pt data vector contains 457 points for DES-Y1 and 533 points for DES-Y3.  

\subsubsection{Systematics in galaxy clustering and weak lensing}\label{sec: Systematics in galaxy clustering and weak lensing}
In this section, we summarize the systematics modelling. We mainly follow the DES-Y3 key projects and point out the difference between DES-Y1 and DES-Y3~\cite{DES:2017tss,DES:2021rex}.
\\ \\ 
\textbf{Galaxy Bias}: The linear galaxy bias is parameterized by a scalar for each redshift bin, i.e. $b^{i}(k,z) = b^i$, for five redshift bins. They are marginalized by a conservative prior $\mathcal{U}(0.8, 3.0)$. We do not consider non-linear galaxy bias in our analysis.
\\ \\
\textbf{Intrinsic Alignment of Galaxy}: In our analysis, we adopt NLA for DES-Y1 and TATT for DES-Y3; their respective parameters are shown in Tables~\ref{table:prior_choices_Y1} and~\ref{table:prior_choices_Y3}. We fix the pivot redshift at $z_0 = 0.62$.
\\ \\
\textbf{Multiplicative shear calibration}: we model the shear calibration with a marginalized parameter $m^i$ for each redshift bin, as shown below:
\begin{equation}
\label{eq:Multiplicative_shear_calibration}
\begin{array}{rlr}
\xi_{\pm}^{i j}(\theta) & \longrightarrow & \left(1+m^{i}\right)\left(1+m^{j}\right) \xi_{\pm}^{i j}(\theta), \\
\gamma^{ij}_t(\theta) & \longrightarrow & \left(1+m^{j}\right) \gamma^{ij}_t(\theta).
\end{array}
\end{equation}
DES-Y1 and DES-Y3 have different calibrations from simulations, detailed in Tables \ref{table:prior_choices_Y1} and \ref{table:prior_choices_Y3}.
\\ \\
\textbf{Photometric redshift uncertainties}:
we model the uncertainties in photometric redshift distribution for both source and lens galaxies by a shift parameter, $\Delta z_{x}^{i}$, unique to each redshift bin $i$, as shown below:
\begin{align}\label{def: photo-z error delta z}
n_{x}^{i}(z)=\hat{n}_{x}^{i}\left(z-\Delta z_{x}^{i}\right), \quad x \in\{\text{source}, \text{lens}\}.
\end{align}
DES-Y1 priors for $\Delta z_{x}^{i}$ differ from DES-Y3 priors and both are shown in Tables~\ref{table:prior_choices_Y1} and~\ref{table:prior_choices_Y3}. We do not model stretches in the photometric redshift distribution of lens galaxies by an additional free parameter $\sigma_z^{i}$ as in the DES-Y3 key project.
\\ \\ 
\textbf{Lensing magnification}: As detailed in \cite{DES:2021rex, 2022arXiv220909782E}, a parameter $C_l^{i}$ is defined to describe the foreground mass effects on the observed number density of lens galaxies. \textbf{The expression that modifies Eq.~\ref{eq: Window of galaxy} can be found in \cite{Fang:2019xat}.} The parameter is calibrated from data for each redshift bin and held fixed in our analysis as shown in Table \ref{table:prior_choices_Y3}. This systematic is not considered for the DES-Y1 data set in this paper.
\\ \\
\textbf{Non-local effects in galaxy-galaxy lensing}: for DES-Y3 specifically, we follow the marginalization approach developed in \citet{MacCrann:2019ntb} and we adopt an informative prior of the point-mass parameter $B_i \in \mathrm{Flat}(-5, 5 )$. Such systematic is not considered for the DES-Y1 data set in this paper.
\\ \\
\textbf{$X_{\text{lens}}$ factor}: A non-physical parameter $X_{\text{lens}}$ was proposed in DES-Y3 to solve the internal inconsistency between galaxy-galaxy lensing and galaxy clustering 2PCF~\cite{DES:2021wwk}. The two lensing samples, \textsc{redMaGiC} and \textsc{MagLim}, show discrepancies between the galaxy bias inferred from galaxy-galaxy lensing and galaxy clustering in the DES-Y3 analysis~\cite{DES:2021zxv,DES:2022urg}. In this work, we adopt \textsc{redMaGiC} lensing sample with fixed $X_{\text{lens}} = 1$ for both DES-Y1 and DES-Y3. We plan to follow up this work with a detailed comparison between \textsc{redMaGiC} and \textsc{MagLim}, including marginalization over $X_{\text{lens}}$ parameter and recent changes to the \textsc{redMaGiC} color selection algorithm~\cite{DES:2021zxv}.

\subsection{Priors and External Data}
\label{sec:data_set_for_geometry}

The split between growth and geometry information is not unique. Within our choices, we select external probes so that DES-Y3 is the only constraining data set on growth parameters besides the boundaries of validity of the Euclid Emulator. We don't present DES-only chains as in~\cite{DES:2020iqt}, since they have shown that DES needs to be combined with external data to provide useful constraining power on the difference between geometry and growth parameters. We combine DES with external data described below:
\\ \\
\textbf{CMBP}: Planck 2018 low-$\ell$ EE polarization data ($\ell$ < 30) in combination with the high-$\ell$ TTTEEE spectra truncated right after the first peak ($35<\ell<396$). Our choice removes late-time Integrated Sachs Wolfe information. It also removes CMB lensing effects as the CMB lensing-induced smoothing on the temperature power spectrum only affects constraints on cosmological parameters when including higher acoustic peaks. We find this prior complementary to the compressed CMB likelihood adopted~\cite{DES:2020iqt}. Our CMB choices are slightly more conservative on $n_{\mathrm{s}}$ and $\Omega_\mathrm{b} h^2$, but they do constrain the early-Universe inflationary amplitude $A_\mathrm{s}$.
\\ \\
\textbf{SNIa}: Pantheon Type Ia supernovae sample~\cite{Pan-STARRS1:2017jku}. Type Ia supernovae are a constraint on geometry parameters only; their likelihood does not require knowledge of the large-scale structure. There are, however, lensing magnification effects on the Hubble diagram~\cite{Wang:2004ax,Zhai:2019sgi,Zhai:2020axw} and growth effects on SNIa peculiar velocity distribution~\cite{Castro:2015rrx,Garcia:2020qah} that will need to be taken into account in future Stage IV surveys; for now, we disregard modeling these growth effects. Note that we do not use the local measurement as the prior on $H_0$ is strongly in tension with other geometry probes~\cite{sh0es21}. 
\\ \\
\textbf{BBN}: We use derived constraint on baryon density from astrophysical probe: $100 \Omega_\mathrm{b}h^2 = 2.208\pm0.052$~\cite{DES:2017txv}. 
\\ \\
\textbf{BAO}: We use Baryon Acoustic Oscillation data from the SDSS DR7 main galaxy sample \cite{Ross:2014qpa} in combination with the 6dF galaxy survey \cite{beutler20116df} at $z_{\text{eff}} = 0.15$ and $z_{\text{eff}} = 0.106$ respectively, and the SDSS BOSS DR12 low-z and CMASS combined galaxy samples at $z = 0.38, 0.51, 0.61$ \cite{BOSS:2016wmc}. These constraints come from comparing the observed scale of the BAO feature and the sound horizon. As a distance measurement of the late universe, we consider BAO to be pure geometry information. 

\begin{figure*}[t]
\centering
\includegraphics[width=0.4\columnwidth]{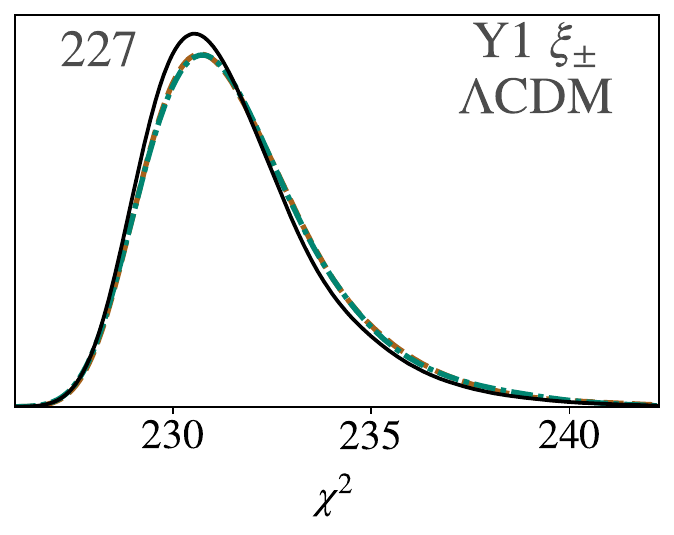} 
\includegraphics[width=0.41\columnwidth]{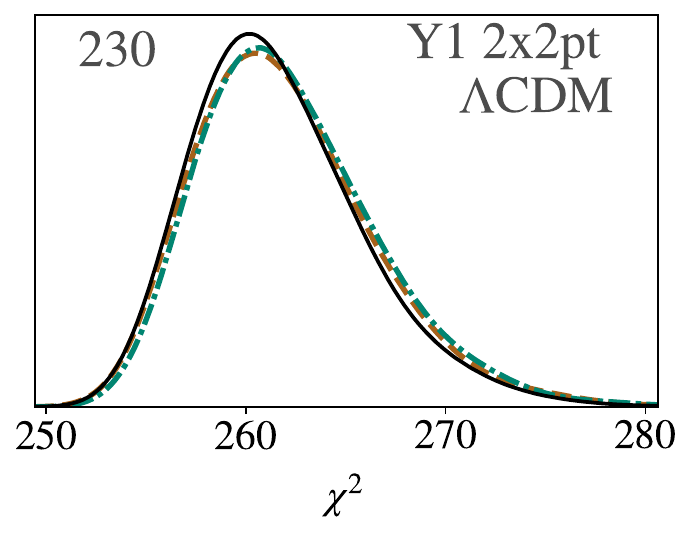}
\includegraphics[width=0.41\columnwidth]{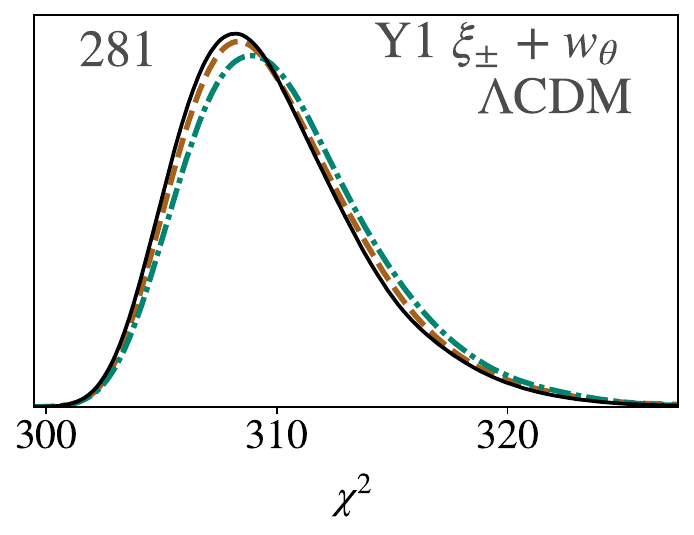}
\includegraphics[width=0.4\columnwidth]{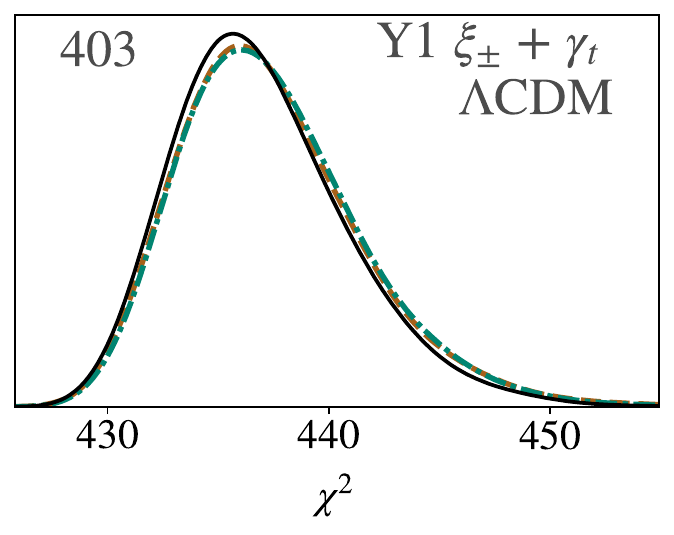} 
\includegraphics[width=0.405\columnwidth]{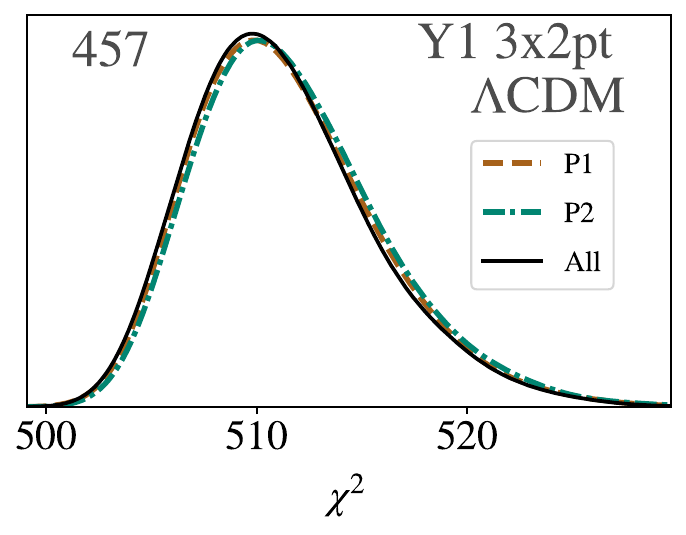}  \\ 
\includegraphics[width=0.4\columnwidth]{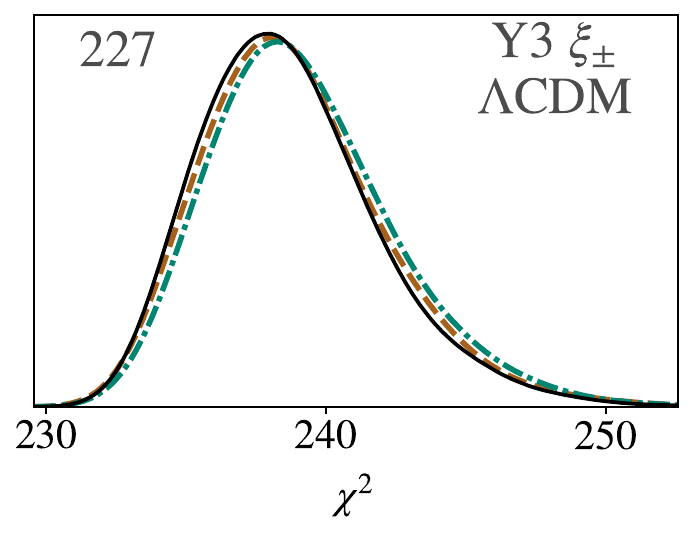}
\includegraphics[width=0.41\columnwidth]{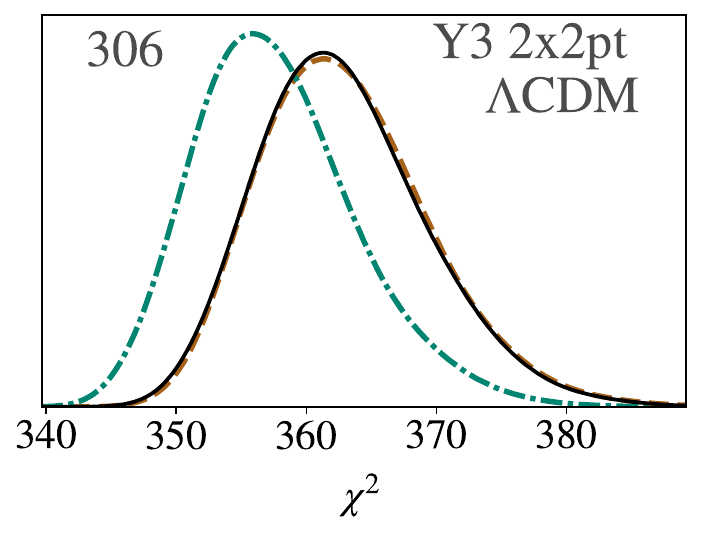} 
\includegraphics[width=0.395\columnwidth]{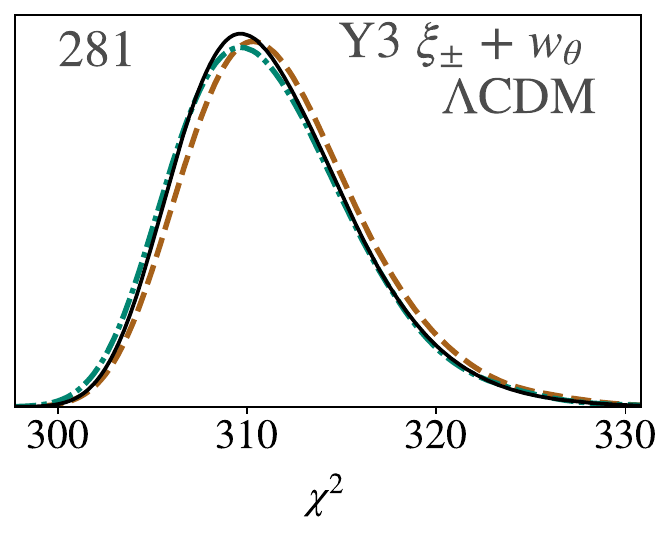}
\includegraphics[width=0.41\columnwidth]{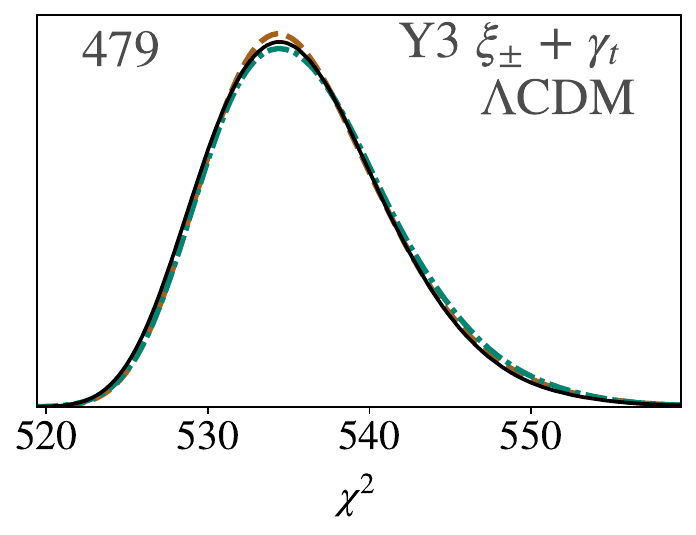}
\includegraphics[width=0.393\columnwidth]{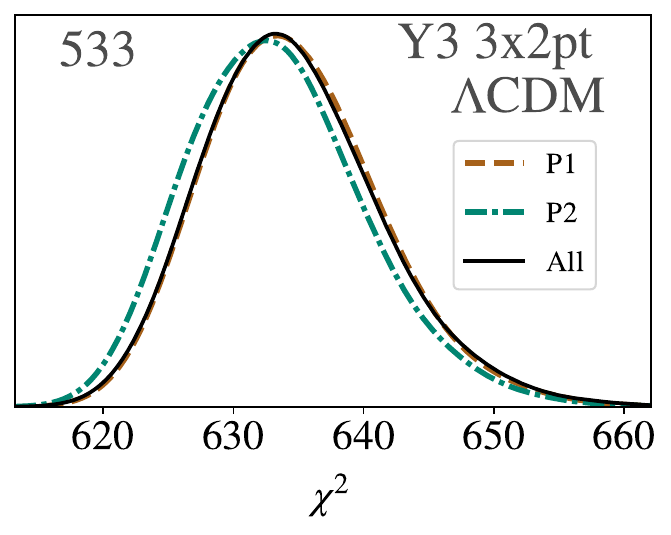} 
\caption{The DES $\chi^2$ distribution of different combinations of two-point correlation functions in split-$\Lambda$CDM chains. The number of data points is printed in the upper left of each panel, after masking us applied for DES-Y1 and DES-Y3, respectively. This plot demonstrates that the P1, P2 and All data priors and external data combinations do not degrade the DES fit except for the DES-Y3 2x2pt. This anomalous data vector combination predicts a small inflationary amplitude, $A_\mathrm{s}$, in $\Lambda$CDM, incompatible with CMB data~\cite{DES:2021wwk}. The cosmic shear cross-correlation reduces this problem considerably on the 3x2pt fit. However, the detailed comparison between $\xi_\pm + w_\theta$ and $\xi_\pm + \gamma_t$ against 3x2pt stands out. Both $\xi_\pm + w_\theta$ and $\xi_\pm + \gamma_t$  combinations show virtually no $\chi^2$ changes between all three priors; the same is not true for 3x2pt.}
\label{fig:210chi2plot}
\end{figure*}

To better understand the effects of these external data sets on the final results, we adopt the following three sets of priors: \\ \\
\textbf{Prior 1 (P1)}: Emulator prior + CMBP \\ \\
\textbf{Prior 2 (P2)}: Emulator prior + SNIa + BAO + BBN \\ \\
\textbf{Prior 3 (All)}: Emulator prior + CMBP + SNIa + BAO + BBN \\ \\
Table~\ref{table:prior_choices_cosmology} summarizes our adopted informative priors on the cosmological parameters. Figure~\ref{fig:prior test1} compares DES-Y1/Y3-only chains with the uninformative priors adopted by the DES collaboration against our P1 and P2 priors. This figure assumes a $\Lambda$CDM model and Halofit for the non-linear matter power spectrum. Our priors are consistent with DES-only posteriors in all parameters, including $\sigma_8$. Since the SNIa + BAO + BBN combination does not provide any information on inflationary parameters, 
the only limits on $A_{\mathrm{s}}$ and $n_{\mathrm{s}}$ in P2 comes from the Euclid Emulator bounds.
Therefore, comparing DES + P1 against DES + P2 chains offers valuable information on how internal DES tensions that shift $A_{\mathrm{s}}$ and $n_{\mathrm{s}}$ affect our results on growth parameters.

\begin{figure*}[t]
\centering
\includegraphics[width=0.4\columnwidth]{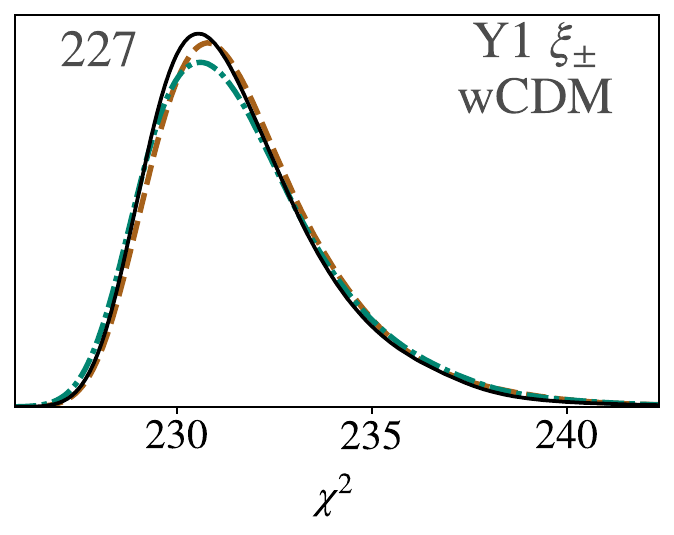}
\includegraphics[width=0.41\columnwidth]{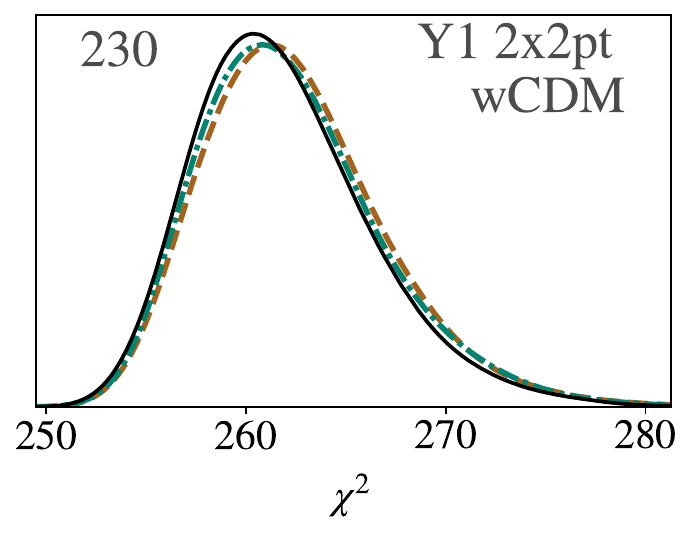} 
\includegraphics[width=0.41\columnwidth]{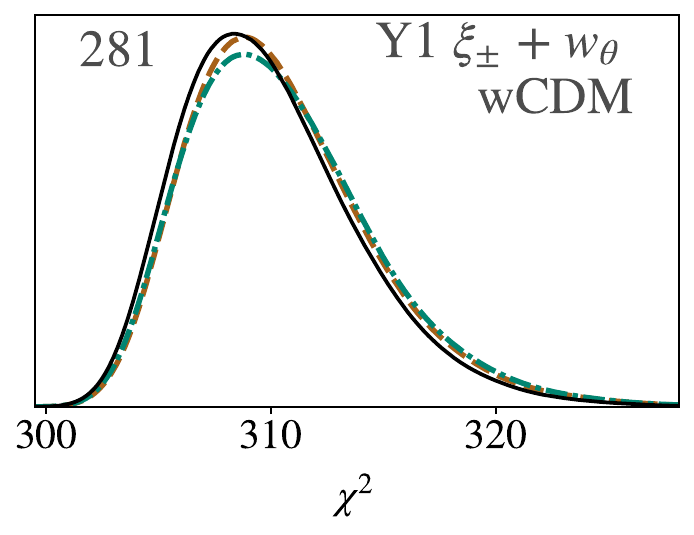}
\includegraphics[width=0.4\columnwidth]{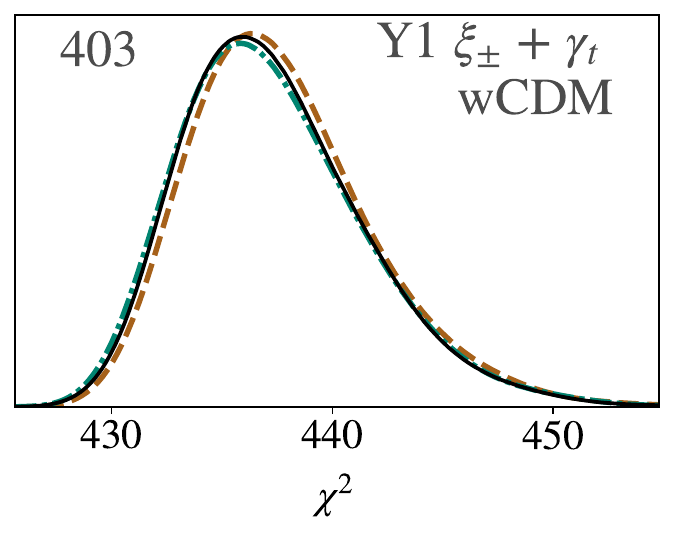}
\includegraphics[width=0.41\columnwidth]{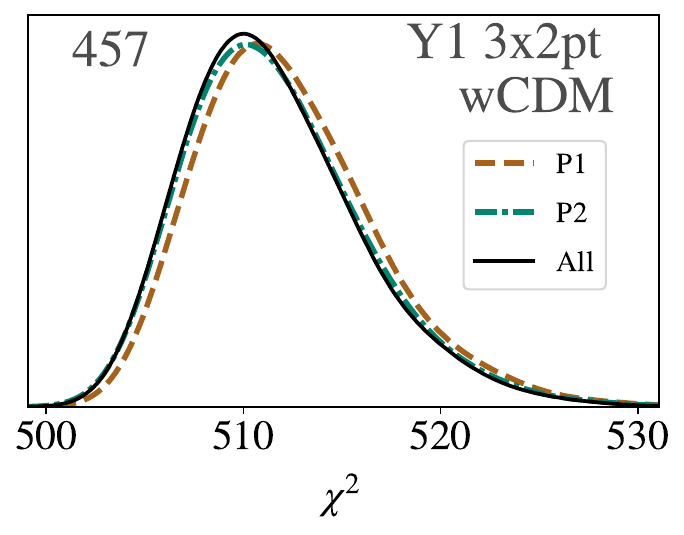} \\
\includegraphics[width=0.4\columnwidth]{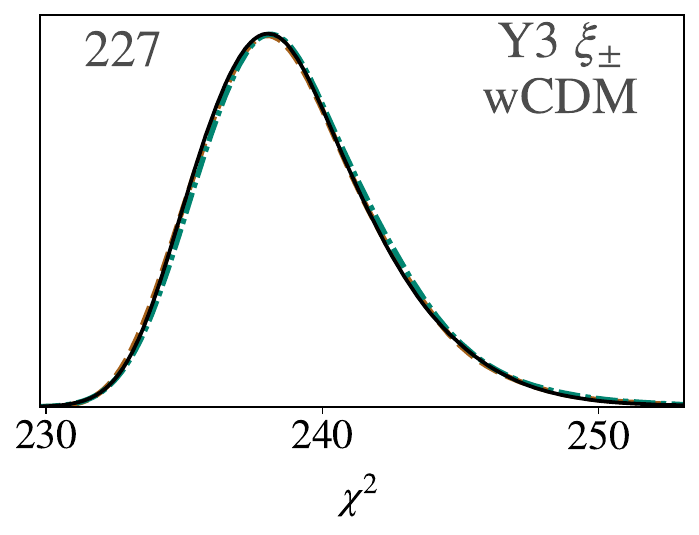}
\includegraphics[width=0.41\columnwidth]{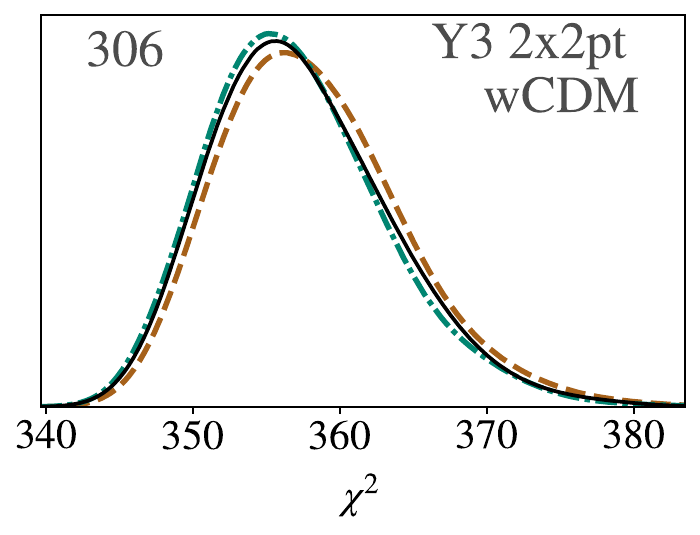} 
\includegraphics[width=0.4\columnwidth]{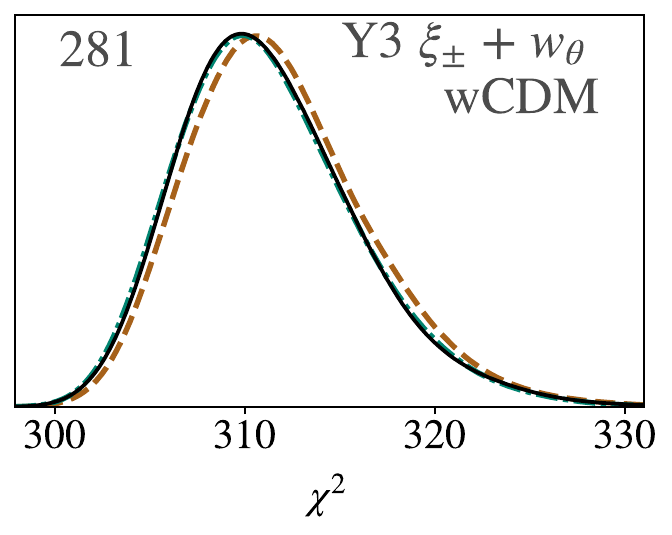}
\includegraphics[width=0.4\columnwidth]{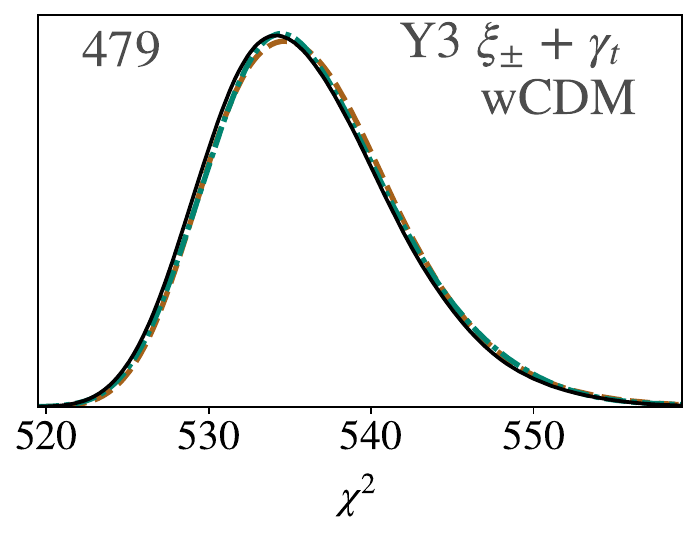}
\includegraphics[width=0.4\columnwidth]{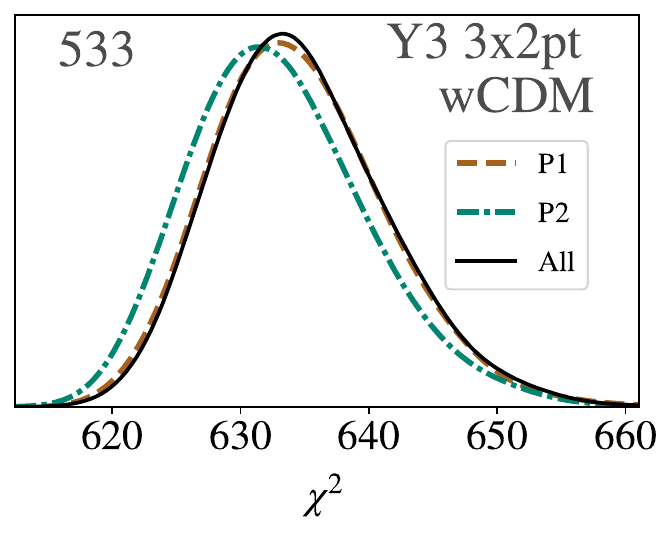}
\caption{The DES $\chi^2$ distribution of different combinations of two-point correlation functions in split-$w$CDM chains. The number of data points is printed in the upper left of each panel, after masking us applied for DES-Y1 and DES-Y3, respectively. This plot demonstrates that the P1, P2 and All data priors and external data combinations do not degrade the DES fit. The 2x2pt data vector anomalous data vector combination prefers a small inflationary amplitude, $A_\mathrm{s}$ in the absence of CMB external data, see Fig.~\ref{fig:1d_result_wCDM}. However, growth parameters can restore the DES 2x2pt goodness-of-fit when $A_\mathrm{s}$ is set by the CMB prior, unlike what we observe in the $\Lambda$CDM split, see Fig.~\ref{fig:210chi2plot}. Indeed, Figure~\ref{fig:1d_result_wCDM} shows that DES-Y3 2x2pt data has higher detection of $\Delta \Omega_\mathrm{m} - \Delta w < 0$ when combined with CMB data. Unfortunately, both $\Lambda$CDM and w-CDM splits have similar goodness-of-fit on the 3x2pt chains that incorporate cosmic shear, including the slight loss of fit when combining DES and CMB data, and $\Delta \Omega_\mathrm{m} - \Delta w$ is consistent with zero.}
\label{fig:210chi2plot_wcdm}
\end{figure*}

Figures~\ref{fig:210chi2plot} and~\ref{fig:210chi2plot_wcdm} show that the DES-Y1 and DES-Y3 $\chi^2$ distributions are nearly independent of prior P1/P2/\textsc{All} choices in both split $\Lambda$CDM and $w$CDM models. The priors are broad enough not to impact the model's DES $\chi^2$ fit, except for the DES-Y3 2x2pt in $\Lambda$CDM split. As we will see, the internal tensions on DES-Y3 2x2pt shift the inflationary parameters to values inconsistent with the CMB prior in both $\Lambda$CDM and $w$CDM splits. In $\Lambda$CDM, $\Delta \Omega_\mathrm{m}$ cannot restore the goodness-of-fit; there is a $\Delta \chi^2 \approx 5$ difference between DES-Y3 2x2pt + P1 and DES-Y3 2x2pt + P2. Interestingly, $\Delta \Omega_\mathrm{m}$ and $\Delta w$ can correct DES-Y3 2x2pt + P1 fit in $w$CDM split.  

\subsection{Pipeline}
\label{sec:pipeline}

We perform the MCMC analysis using \textsc{Cocoa}, the \textsc{Cobaya}-\textsc{Cosmolike} Architecture~\cite{cocoa}. \textsc{Cocoa} is a modified version of \textsc{CosmoLike}~\cite{Krause:2016jvl} multi-probe analysis software incorporated into the \textsc{Cobaya} framework~\cite{Torrado:2020dgo}. DES-Y1 and DES-Y3 covariance matrices were computed using \textsc{CosmoCov}~\cite{Fang:2020vhc}. \textsc{CosmoCov} and \textsc{Cocoa} are both derived from \textsc{Cosmolike}~\cite{Krause:2016jvl}, the former pipeline computes covariance matrices, and the latter evaluates data vectors. \textsc{CosmoLike} within \textsc{Cocoa} has efficient OpenMP shared-memory parallelization~\cite{openmp} and cache system compatible with the slow-fast decomposition implemented in the default \textsc{Cobaya} Monte-Carlo Markov chain sampler (MCMC) The OpenMP efficiency in \textsc{CosmoLike} is around $50\%$, i.e., quadrupling the number of OpenMP cores halves \textsc{CosmoLike} runtime. 

\textsc{CosmoLike} has been used in both DES-Y1/Y3 multi-probe analyses when constraining $\Lambda$CDM parameters~\cite{DES:2017tss,DES:2021rex} and for calibrating Bayesian evidences~\cite{Miranda:2020lpk}. It has also been used in forecast studies for Rubin Observatory's LSST and Roman Space Telescope~\cite{Eifler:2013fit,Eifler:2020vvg,Eifler:2020hoy, DESC-SRD}. 

We compute the linear power spectrum with the \textsc{CAMB} Boltzmann code~\cite{Lewis:2002ah, Howlett_2012};  \textsc{Cobaya} already had implementations of all external data sets. We adopt \textsc{Cobaya}'s default adaptive metropolis hasting MCMC sampler, and we employ the Gelman-Rubin criteria $R-1 < 0.02$ to establish chain convergence~\cite{10.1214/ss/1177011136}. We post-process chains and creat figures using \textsc{GetDist}~\cite{getdist}. 

Changes in the growth parameters are only semi-fast; they do not require CAMB to recompute distances and matter power spectrum; only \textsc{CosmoLike} must be rerun to update the DES data vectors. Due to an efficient cache system, \textsc{CosmoLike} reruns with only modified growth parameters takes about half the runtime compared with when all parameters are varied. CAMB and \textsc{CosmoLike} runtimes are roughly equal; the time ratio between slow and semi-fast parameters is, therefore, approximately 4:1. The 3x2pt data vector evaluation time with 10 OpenMP cores is of order 1.5s on modern \textsc{AMD EPYC 7642} 48-core nodes. This estimation includes CAMB evaluation, non-limber integration, and TATT modeling. Finally, code comparisons between the \textsc{CosmoSiS} pipeline and \textsc{Cosmolike} were presented in ~\cite{DES:2017tss,DES:2021rex}.

\subsection{Validation on Synthetic Data}
\label{sec:validation}

In this section,  we generate a synthetic noiseless $\Lambda$CDM data vector from \textit{Planck} best fit cosmological parameters without lensing: $\big\{A_{\mathrm{s}}\!\!\times\!\!10^{-9}, n_{\mathrm{s}}, H_0, \Omega_{\rm m}, \Omega_{\rm b}\big\} = \big\{2.101, 0.965, 67.32, 0.317, 0.049 \big\}$. This set of parameters is compatible with both P1 and P2 priors~\cite{Planck:2018nkj}. We run MCMCs, including all nuisance parameters, and see if the posterior would give equal growth and geometry parameters at the fiducial value.

\begin{figure}
\includegraphics[width=0.95\columnwidth]{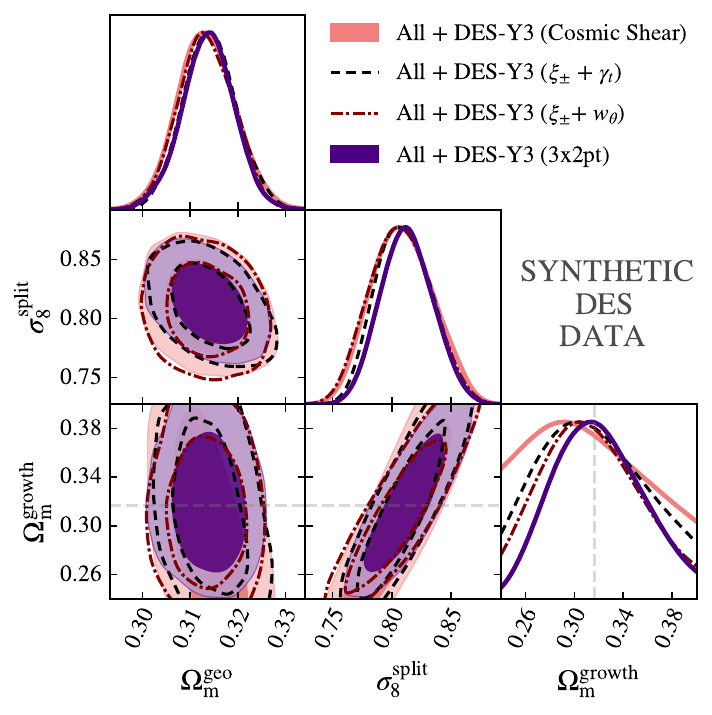}
\caption{Posteriors derived from different combinations of synthetic DES-Y3 2PCFs in the split $\Lambda$CDM model. As described in Sec.~\ref{sec:data_set_for_geometry}, the \textsc{All} external data combination consists of CMBP + SNIa + BAO + BBN, with CMBP being Planck 2018 low-$\ell$ EE polarization data and the high-$\ell$ TTTEEE spectra truncated right after the first peak ($35<\ell<396$). Prior to the growth parameter $0.24 \leq \Omega_\mathrm{m}^\mathrm{growth} \leq 0.4$ is compatible with the Euclid Emulator boundaries. All posteriors are prior limited, but the plot clarifies the gain in constraining power when galaxy-galaxy lensing and galaxy-clustering are added to cosmic shear. }
\label{fig:plot36_S8}
\end{figure}

\subsubsection{Comparison between Cosmic Shear and 3x2pt}
\label{sec:comparison_between_CS_and_3pt2}

Assuming the \textsc{All} prior and external data combination, there is a significant improvement on $\Omega_\mathrm{m}^{\rm growth}$ constraints in $\Lambda$CDM split when going from cosmic shear to 3x2pt, as shown in Fig.~\ref{fig:plot36_S8}. In the 3x2pt case, the $\Omega_\mathrm{m}^{\rm growth}$  posterior is well centered at the fiducial value, while cosmic shear provides only marginal improvements compared with the uniform $0.24 < \Omega_\mathrm{m}^{\text{\rm growth}} < 0.4$ prior. This narrow prior is informative in both chains, the boundary coming from the range of the~\textsc{Euclid Emulator}. Improving the small-scale modeling validity of cosmic shear and 2x2pt may tighten the $95\%$ confidence level of $\Omega_\mathrm{m}^{\rm growth}$ enough to be within the allowed range; ~\cite{hem20} and~\cite{2022PhRvD.106d3520P} offer a roadmap on how to implement such improvements in future work.

\begin{figure}[t]
\includegraphics[width=0.95\columnwidth]{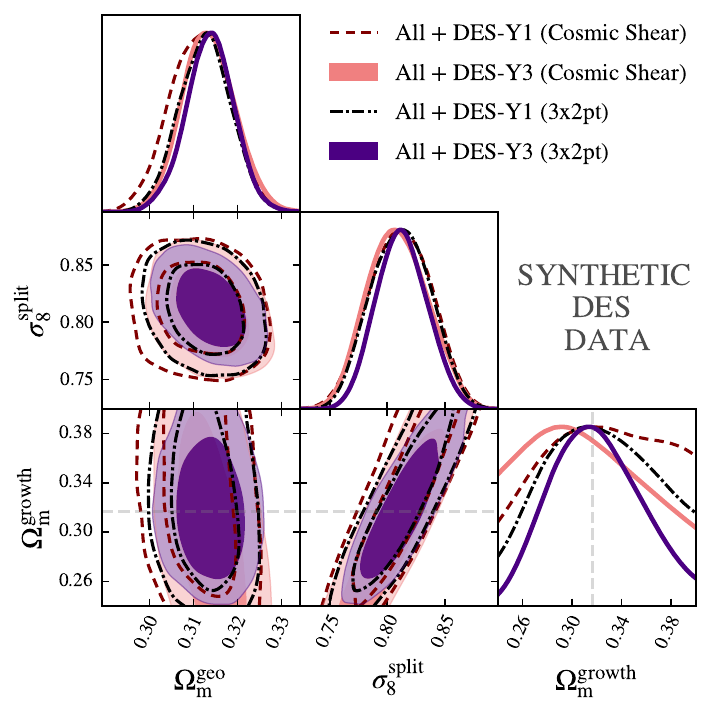}
\caption{Split $\Lambda$CDM posteriors derived from cosmic shear and 3x2pt combined with the \textsc{All} external data combination. As described in Sec.~\ref{sec:data_set_for_geometry}, the \textsc{All} external data combination consists of CMBP + SNIa + BAO + BBN, with CMBP being Planck 2018 low-$\ell$ EE polarization data and the high-$\ell$ TTTEEE spectra truncated right after the first peak ($35<\ell<396$). DES-Y1 3x2pt error bar on $\Omega_\mathrm{m}^\mathrm{growth}$ is approximately $17\% $ larger compared with DES-Y3. On the other hand, cosmic shear DES-Y1 and DES-Y3 constraints are similar, and both are prior dominated.}
\label{fig:plot37_S8}
\end{figure}

DES 3x2pt combinations with \textsc{P1} and \textsc{All} external data show nearly identical constraining power on $\Omega_\mathrm{m}^{\rm growth}$. Combined priors on the primordial power spectrum (amplitude and shift) and the shape parameter $\Gamma \equiv \Omega_{\rm m} h$ are the needed external information so that DES can tightly measure growth. The CMBP data alone provides both information while the SNIa + BAO + BBN measurements on $\Omega_{\rm m}^\mathrm{geo}$ and $H_0$ only restrict the shape parameter. Adding more CMB multipoles would improve constraints on early-Universe parameters even more. However, CMB temperature and polarization power spectra are more sensitive to lensing effects on smaller scales, which would limit our ability to compare DES effects on growth parameters against CMB lensing. Partial delensing can alleviate this limitation~\cite{2017JCAP...05..035C}. Another possibility is to consider all multipoles up to $\ell_{\text{max}} \approx 1600$ where effects from nonlinear dark matter collapse are negligible. In this case, however, we would marginalize the chains over lensing principal components so that there is no leakage of information on growth parameters~\cite{Motloch:2018pjy}.   

In the $w$CDM split model, $\Omega_\mathrm{m}^{\rm growth}$ and $w^{\rm growth}$ are not well constrained even in the most informative 3x2pt case. We then show real data constraints on the principal component combination
\begin{equation} 
\label{eq:pc1}
\mathrm{PC}_1 = -0.7071 \Delta w + 0.7071 \Delta \Omega_\mathrm{m} . 
\end{equation}
In both cosmic shear and 3x2pt chains, $\mathrm{PC}_1$ constraints are prior dominated but well centered around zero.

\subsubsection{Comparison between DES-Y1 and DES-Y3}
\label{sec:comparison_between_Y1_and_Y3}

In all three combinations with external data, posteriors on $\Omega_\mathrm{m}^{\rm growth}$ in the $\Lambda$CDM split from DES-Y1 and DES-Y3 cosmic shear are similar, despite the additional nuisance parameters introduced by the TATT intrinsic alignment model in DES-Y3 (see Fig.~\ref{fig:plot37_S8}). We have yet to check if we can obtain more constraining power on growth parameters by adopting the more straightforward NLA model on DES-Y3. On the other hand, the error bar on $\Omega_\mathrm{m}^{\rm growth}$ derived from 3x2pt combined with the \textsc{All} prior is $17\%$ larger in DES-Y1. One caveat to this result is that we have not tested whether expanding the adopted priors on point mass marginalization to the more conservative range $\mathrm{Flat}(-100, 100)$ would significantly degrade DES-Y3 constraints.

The first principle component $\mathrm{PC}_1$, defined on Eq.~\ref{eq:pc1}, has nearly identical and prior dominated DES-Y1 and DES-Y3 posteriors in the $w$CDM split. Additional information from either smaller scales in the 3x2pt data vector or external growth information from CMB lensing and RSD are potential opportunities in future analyses. Figure~\ref{fig:sigma8} shows that $\Omega_\mathrm{m}^{\rm growth}$ and $w^{\rm growth}$ induce changes on $\sigma_8^{\rm split}(z)$ with different redshift evolution. Including high redshift $z>1$ lensing samples from the future Roman Space Telescope may therefore be the key to disentangling growth parameters in $w$CDM split~\cite{Eifler:2020hoy}.

There are near-future possibilities that may expand the redshift range adopted in this paper. \textsc{RedMaGiC} fifth bin, with range $0.8 < z < 0.9$, shows large $X_\mathrm{lens}$ biases~\cite{2022PhRvD.106d3520P}. The alternative DES-Y3 \textsc{MagLim} sample of lens galaxies does have an additional redshift bin in the range $0.95 < z < 1.05$ not accessible by \textsc{redMaGiC}~\cite{Y3MagLim}. However, \textsc{MagLim} high redshift bins were not adopted in the 3x2pt analysis by the DES collaboration and may require further studies on the presence of potential systematic biases~\cite{DES:2021wwk}. Finally, there is the emergent idea of using the same galaxy sample for both clustering and lensing that could potentially expand DES-Y3 constraints on $\sigma_8^{\rm split}(z)$ beyond $z>1$~\cite{Fang:2021ici}.

\section{Results}\label{sec:results}
We split our results section into three components:  starting with a discussion of our results in the $\Lambda$CDM parameter space, we then move to the $w$CDM space, and conclude with quantifying tensions between different probe combinations in the context of both parameter spaces.

\begin{figure}[h!]
\centering
\includegraphics[width=\columnwidth]{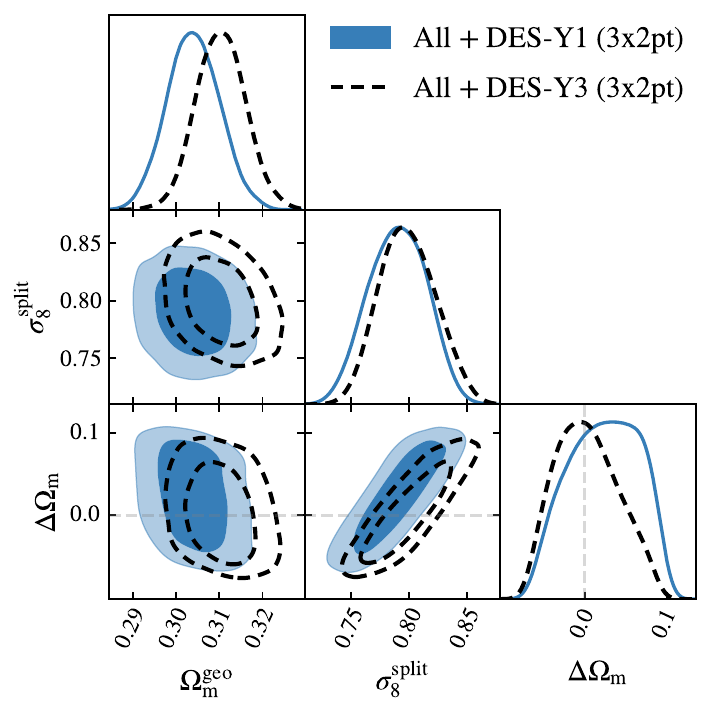}
\caption{Split $\Lambda$CDM posteriors derived from 3x2pt DES-Y1 and DES-Y3  data. As described in Sec.~\ref{sec:data_set_for_geometry}, the All external data combination consists of CMBP + SNIa + BAO + BBN, with CMBP being Planck 2018 low-$\ell$ EE polarization data and the high-$\ell$ TTTEEE spectra truncated right after the first peak ($35<\ell<396$). DES-Y1 3x2pt error bar on growth dark matter density is approximately $10\% $ larger compared with DES-Y3.}. 
\label{fig:204_208_3x2pt} 
\end{figure}

\begin{figure*}[t]
\centering
\includegraphics[width=\columnwidth]{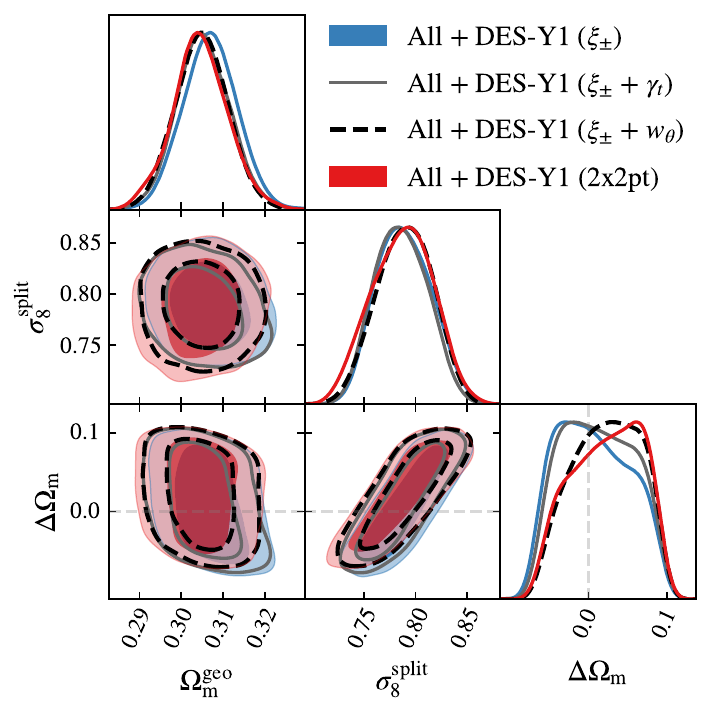}
\includegraphics[width=\columnwidth]{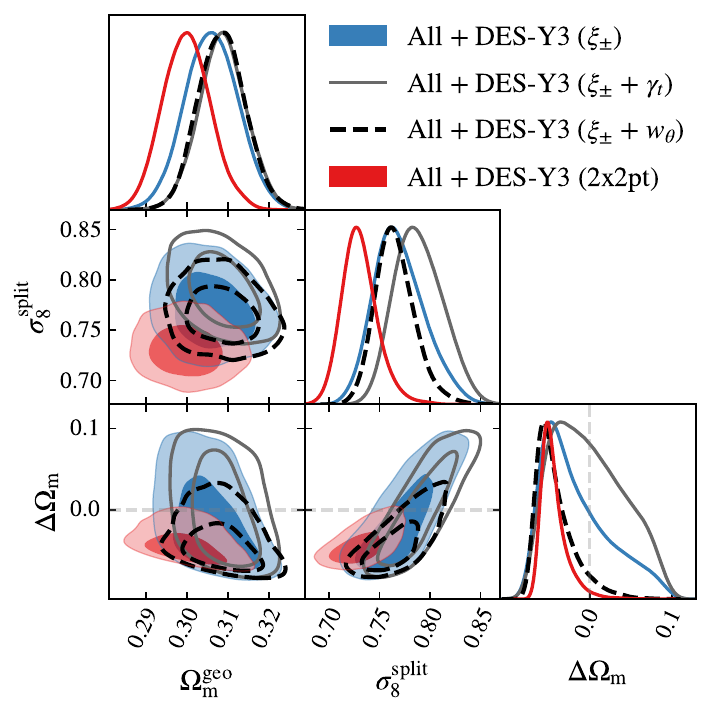}
\caption{Split $\Lambda$CDM posteriors derived from multiple 2PCF combinations in DES-Y1 (left panel) and DES-Y3 (right panel). As described in Sec.~\ref{sec:data_set_for_geometry}, the All external data combination consists of CMBP + SNIa + BAO + BBN, with CMBP being Planck 2018 low-$\ell$ EE polarization data and the high-$\ell$ TTTEEE spectra truncated right after the first peak ($35<\ell<396$). Right panel shows that the DES-Y3 $\xi_\pm + \gamma_t$, $\xi_\pm + w(\theta)$ and 2x2pt all prefer lower values for the $\Omega_\mathrm{m}^\mathrm{growth}$ with upper limits at 95\% confidence level being $0.375$, $0.314$ and $0.288$ respectively. We emphasize that the apparent constraints at $\Delta \Omega_\mathrm{m} \equiv \Omega_\mathrm{m}^\mathrm{growth} - \Omega_\mathrm{m}^\mathrm{geo} \approx -0.8$ is due to effective priors.} 
\label{fig:204_208} 
\end{figure*}a

\begin{figure*}[t]
\centering
\includegraphics[width=2\columnwidth]{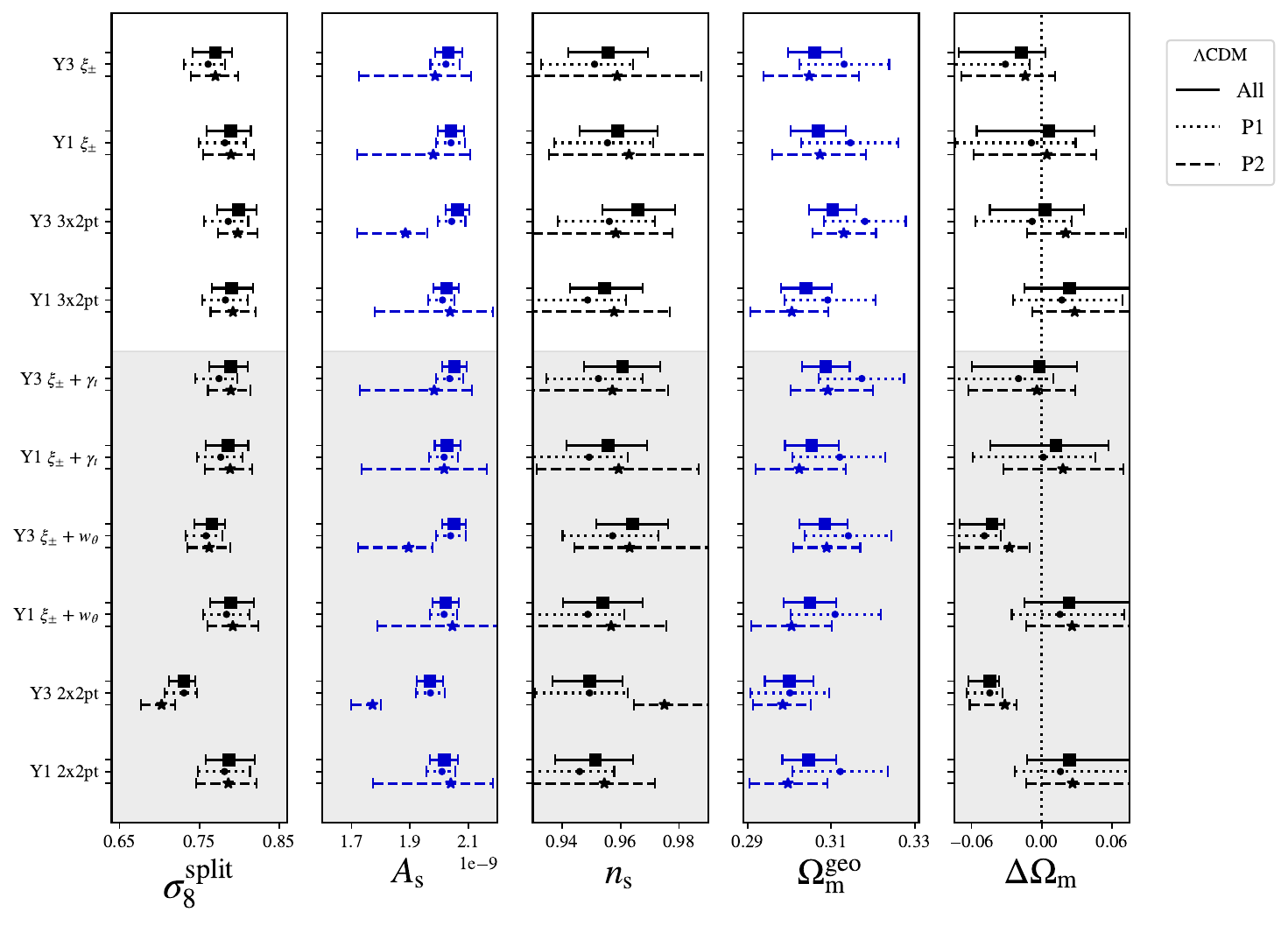}
\caption{One-dimensional posteriors in split $\Lambda$CDM for various DES-Y1 and DES-Y3 two-point correlation functions, with error bars corresponding to marginalized $68\%$ confidence intervals. As described in Sec.~\ref{sec:data_set_for_geometry}, the \textsc{All} external data combination consists of CMBP + SNIa + BAO + BBN, with CMBP being Planck 2018 low-$\ell$ EE polarization data and the high-$\ell$ TTTEEE spectra truncated right after the first peak ($35<\ell<396$). The P1 external data combination is restricted to CMBP, while P2 is SNIa + BAO + BBN. Priors on cosmological parameters are summarized in Table~\ref{table:prior_choices_cosmology}; we define $\Delta\Omega_\mathrm{m} \equiv \Omega_\mathrm{m}^\mathrm{growth} - \Omega_\mathrm{m}^\mathrm{geo}$. The grey background separates our primary results from other probe combinations.}
\label{fig:1d_result_LCDM} 
\end{figure*}

\subsection{Growth-geometry split results in $\Lambda$CDM}
For the most constraining probe combination, DES 3x2pt+\textsc{All}, we show the DES-Y1 and DES-Y3 $\Lambda$CDM results in Fig. \ref{fig:204_208_3x2pt}. In both cases we find no measurable detection of $\Delta \Omega_\mathrm{m}$ being different from zero. This constitutes the main, fiducial result of this paper. 

We explore subsets of the 3x2pt probe combination in Fig.~\ref{fig:204_208}, where the left panel refers to DES-Y1 and the right corresponds to DES-Y3. For DES-Y1 we find that in all cases $\Delta \Omega_\mathrm{m}$ is compatible with zero even within one sigma. For DES-Y3, however, we see shifts from $\Delta \Omega_\mathrm{m}=0$, especially in the 2x2pt (galaxy clustering+galaxy galaxy lensing) case.

We consider this further in Fig.~\ref{fig:1d_result_LCDM}, where we show the one-dimensional posterior distributions on all relevant $\Lambda$CDM split parameters, finding that
except for DES-Y3 $\xi_\pm$+P2, $\xi_\pm + w_\theta$ and 2x2pt chains, all combinations of two-point correlation functions predict $\Delta \Omega_\mathrm{m}$ compatible with zero within one sigma. The deviation on $\xi_\pm$ + P2 is less than two-sigma. Similarly, all combinations between DES 2PCFs and the P2 external data predict $A_\mathrm{s}$ and $n_\mathrm{s}$ values compatible with CMB data on P1/\textsc{All},  except for DES-Y3 $\xi_\pm + w_\theta$ and 2x2pt.

Similarly to what we observe in $\Lambda$CDM chains with synthetic data vectors, DES-Y1 and DES-Y3 cosmic shear provide little information on $\Delta \Omega_\mathrm{m}$ even with the P1/P2/\textsc{All} priors. The additional nuisance parameters introduced by the TATT intrinsic alignment model and point mass marginalization in DES-Y3 do not reduce constraining power for the growth parameters. The situation in the 3x2pt chains is different; the DES-Y1 3x2pt + \textsc{All} error bars are 10\% larger than in DES-Y3, not that far from the predicted 17\% improvement in the synthetic noise-free chains. Priors on $\Omega_\mathrm{m}^\mathrm{growth}$ are still informative, but to a much lesser degree on both DES-Y1 and DES-Y3 3x2pt compared with their cosmic shear counterpart.  

All of the DES-Y1 $\Lambda$CDM split chains are compatible with $\Delta \Omega_\mathrm{m} = 0$; Figs.~\ref{fig:204_208} (left panel) and~\ref{fig:1d_result_LCDM} show large consistency between parameter posteriors derived from all 2PCFs combinations. There are also no appreciable parameter shifts between chains with and without CMB priors; goodness-of-fit is identical in these chains (see Fig.~\ref{fig:210chi2plot}). As expected, $A_\mathrm{s}$ and $n_\mathrm{s}$ constraints are significantly tighter when CMB data is present. Finally, chains that include galaxy clustering (3x2pt, 2x2pt, and $\xi_\pm + w_\theta$) show a small shift towards $\Delta \Omega_\mathrm{m} > 0$, but are still compatible with zero at 68\% confidence level.

For DES-Y3 we see that the $\xi_\pm + w_\theta$ and 2x2pt chains predict, in combination with the \textsc{All} prior, $\Delta\Omega_\mathrm{m} \neq 0$ at 1.75$\sigma$ and 2.60$\sigma$ in statistical significance (see Fig.~\ref{fig:204_208}). We attribute these findings to the well-known incompatibilities between galaxy clustering and galaxy-galaxy lensing in DES-Y3 when using the \textsc{redMaGiC} lens sample.

\subsection{Growth-geometry split results in $w$CDM}
\begin{figure*}[t]
\centering
\includegraphics[width=\columnwidth]{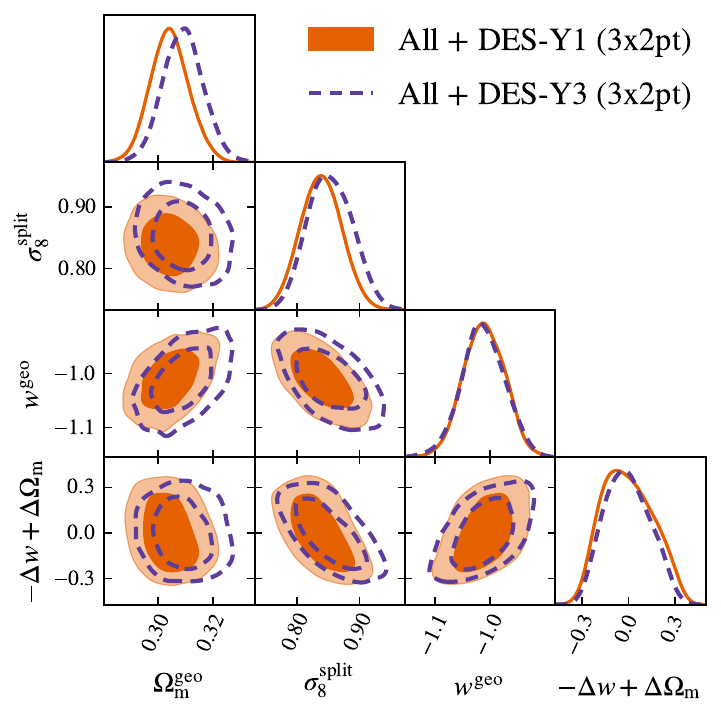}
\includegraphics[width=\columnwidth]{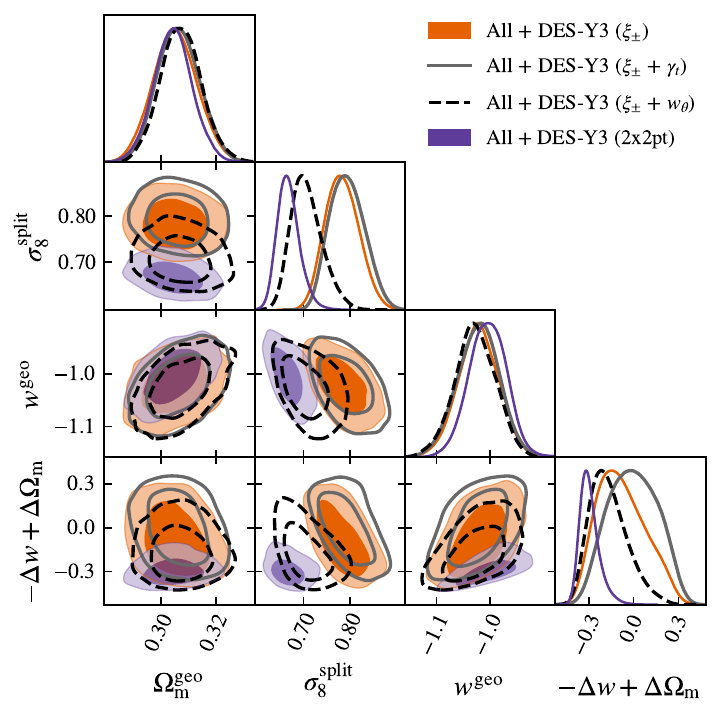}
\caption{Split $w$CDM posteriors derived from 3x2pt (left panel) and multiple 2PCF combinations (right panel). As described in Sec.~\ref{sec:data_set_for_geometry}, the \textsc{All} external data combination consists of CMBP + SNIa + BAO + BBN, with CMBP being Planck 2018 low-$\ell$ EE polarization data and the high-$\ell$ TTTEEE spectra truncated right after the first peak ($35<\ell<396$). Table~\ref{table:prior_choices_cosmology} presents the priors on the cosmological parameters; we define $\Delta\Omega_\mathrm{m} \equiv \Omega_\mathrm{m}^\mathrm{growth} - \Omega_\mathrm{m}^\mathrm{geo}$ and $\Delta w \equiv w^\mathrm{growth} - w^\mathrm{geo}$. All constraints on the combination $ \Delta\Omega_\mathrm{m} - \Delta w$ are prior dominated given the range limitations of $\Omega_\mathrm{m}^\mathrm{growth}$.}
\label{fig:205} 
\end{figure*}

\begin{figure*}[t]
\centering
\includegraphics[width=2.0\columnwidth]{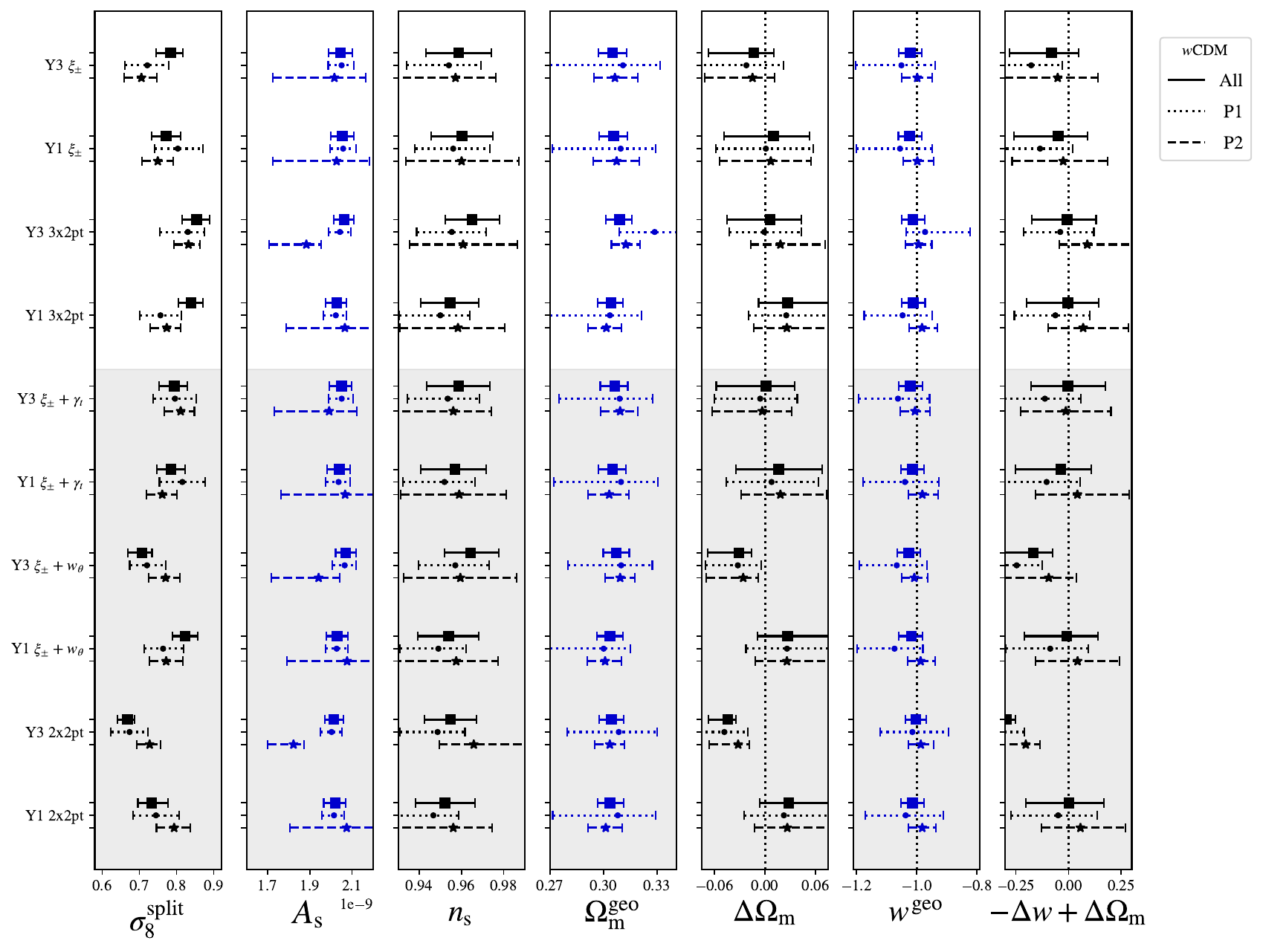}
\caption{One-dimensional posteriors in split $w$CDM for various DES-Y1 and DES-Y3 two-point correlation functions, with error bars corresponding to marginalized $68\%$ confidence intervals. As described in Sec.~\ref{sec:data_set_for_geometry}, the \textsc{All} external data combination consists of CMBP + SNIa + BAO + BBN, with CMBP being Planck 2018 low-$\ell$ EE polarization data and the high-$\ell$ TTTEEE spectra truncated right after the first peak ($35<\ell<396$). The P1 external data combination is restricted to CMBP, while P2 is SNIa + BAO + BBN. Priors on cosmological parameters are summarized in Table~\ref{table:prior_choices_cosmology}; we define $\Delta\Omega_\mathrm{m} \equiv \Omega_\mathrm{m}^\mathrm{growth} - \Omega_\mathrm{m}^\mathrm{geo}$ and $\Delta w \equiv w^\mathrm{growth} - w^\mathrm{geo}$. The grey background separates our primary results from other probe combinations. As could be expected, the DES-Y3 2x2pt + P2 predicts lower values for the inflationary amplitude $A_\mathrm{s}$ incompatible with CMB priors. The DES-Y3 2x2pt also predicts non-zero values for the principal component $ \Delta\Omega_\mathrm{m} - \Delta w$ for all external data combinations; the more extreme deviations being DES-Y3 2x2pt + All with mean $-0.296$ and standard deviation $0.066$.}
\label{fig:1d_result_wCDM}  
\end{figure*}

For the $w$CDM parameter space we summarize our results in Figs.~\ref{fig:205} and \ref{fig:1d_result_wCDM}, where the former again shows selected results in two dimensions and the latter summarizes all chains in one-dimensional projections. Qualitatively, we see similar behaviour as in the $\Lambda$CDM case. While DES cosmic shear and 3x2pt data shows $\Delta\Omega_\mathrm{m} - \Delta w$ being consistent with zero, the picture becomes more complicated when considering subsets of the 3x2pt case that involve galaxy clustering of \textsc{redMaGiC}. 

In particular, the 2x2pt + \textsc{All} chain favors $\Delta\Omega_\mathrm{m} - \Delta w < 0$ at 4.48$\sigma$, higher than any $\Delta \Omega_\mathrm{m} \neq 0$ detection in $\Lambda$CDM split. The $w$CDM split 2x2pt + \textsc{All} chain also predict quite low $\sigma_8^\mathrm{split} = 0.682 \pm 0.0243$, while in $\Lambda$CDM we have $\sigma_8^\mathrm{split} = 0.730 \pm 0.1813$. 

While a 4.48$\sigma$ detection is significant, we again refrain from claiming new physics in the $w$CDM model space, due to the aforementioned problems with the DES-Y3 \textsc{redMaGiC} sample. Instead, we plan to further investigate growth-geometry split with alternative lens samples and when marginalizing over $X_\mathrm{lens}$.

\subsection{Quantifying tensions between probes}
\label{sec:TensionMetric}

\subsubsection{Method}

To evaluate the tension we use the parameter difference method~\cite{raveri_non-gaussian_2021,lemos_assessing_2021}. Given two chains $\theta_1$ and $\theta_2$ and their corresponding posteriors $\mathcal{P}_1(\theta_1)$ and $\mathcal{P}_2(\theta_2)$, begin by computing the difference between these two chains, denoted with $\Delta\theta = \theta_1-\theta_2$. Using this difference chain we can write $\mathcal{P}_2(\theta_2) = \mathcal{P}_2(\theta_1-\Delta\theta)$. By marginalizing over $\theta_1$ we get the parameter difference posterior,
\begin{equation}
    \mathcal{P}(\Delta\theta) = \int\limits_{V_\Pi} \mathcal{P}_1(\theta_1)\mathcal{P}_2(\theta_1-\Delta\theta)d\theta_1 \, ,
\end{equation}
where $V_\Pi$ is the subset of the domain covered by the prior. As $\theta_1\rightarrow\theta_2$, the means of each chain approach equality and the mean of the parameter difference chain approaches 0. Thus the volume of the regions with $\mathcal{P}(\Delta\theta)>\mathcal{P}(0)$ approaches 0, so we can approximate the tension using
\begin{equation}\label{volume_int}
    \Delta = \int\limits_{\mathcal{P}(\Delta\theta)>\mathcal{P}(0)}\mathcal{P}(\Delta\theta)d\Delta\theta \, .
\end{equation}
This volume is interpreted as a probability of parameter shift, denoted $\Delta$. If $\Delta$ comes from a Gaussian distribution, the number of standard deviations from $0$ is given by
\begin{equation}
    n_\sigma = \sqrt{2}\text{Erf}^{-1}(\Delta) \, .
\end{equation}
The resulting $n_\sigma$ is reported.

To estimate the posterior we use Masked Autoregressive Flows (MAFs)~\cite{raveri_non-gaussian_2021,papamakarios_masked_2018}, which is a neural network that learns an invertible mapping from an arbitrary parameter space to a gaussianized one. The loss function for MAFs is the negative log probability from a unit Gaussian. Due to the autoregressive property, the Jacobian is triangular and thus the determinant is tractable to compute even for a large number of dimensions. Thus we can estimate the posterior as a reparameterization of a Gaussian and find the log-probability of arbitrary points.

Before training the neural network, we follow the implementation in ref.~\cite{raveri_non-gaussian_2021} to apply a linear transformation to $\Delta\theta$ given from the Gaussian approximation for $\mathcal{P}(\Delta\theta)$
\begin{equation}
    \Delta\theta' = C^{-1}(\Delta\theta - \mu) \, ,
\end{equation}
with $C$ the covariance and $\mu$ the mean of $\mathcal{P}(\Delta\theta)$, then map $\Delta\theta'$ to the fully Gaussianized parameter space. This enhances the convergence rate of the neural networks. Denoting the learned mapping as $\phi(\Delta\theta') = y$ and the unit Gaussian density as $\mathcal{N}$, we can then relate the log-probability as
\begin{equation}
    \mathcal{P}(\Delta\theta) = \mathcal{N}(y) \frac{|\det(J_\phi (\Delta\theta'))|}{|\det(C)|}
\end{equation}
where $J_\phi$ denotes the Jacobian of $\phi$.

To compute the integral in Eq.~\ref{volume_int} we use Monte Carlo integration. Using the MAF we randomly sample from the posterior and calculate the log probability. The fraction of generated points that land in the region $\mathcal{P}(\Delta\theta)>\mathcal{P}(0)$ are counted. The error of the numerical integration is given by the Clopper-Pearson interval for a binomial distribution.

\subsubsection{Results}
\begin{figure*}[t]
\centering
\includegraphics[width=0.97\columnwidth]{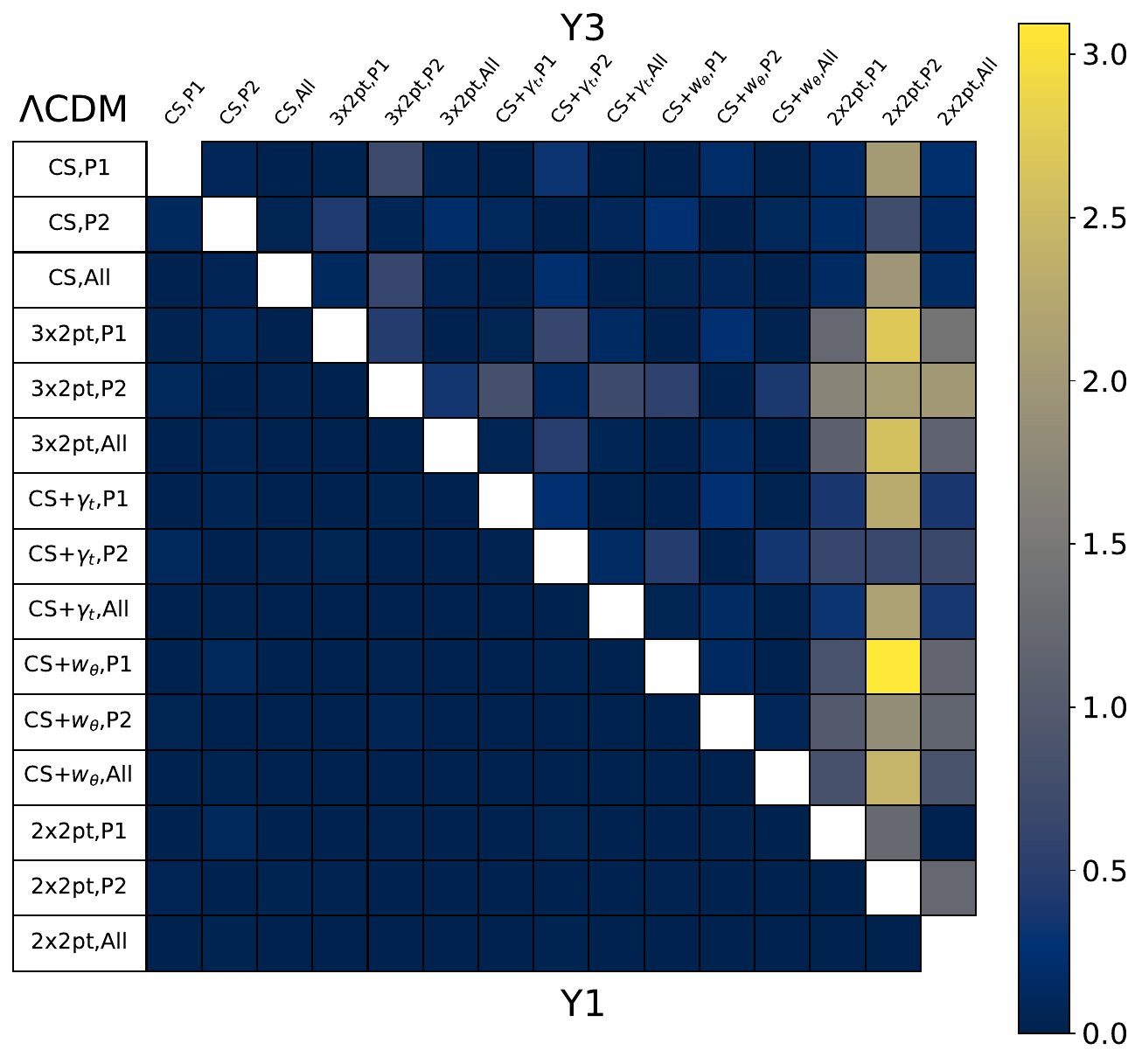}
\includegraphics[width=0.97\columnwidth]{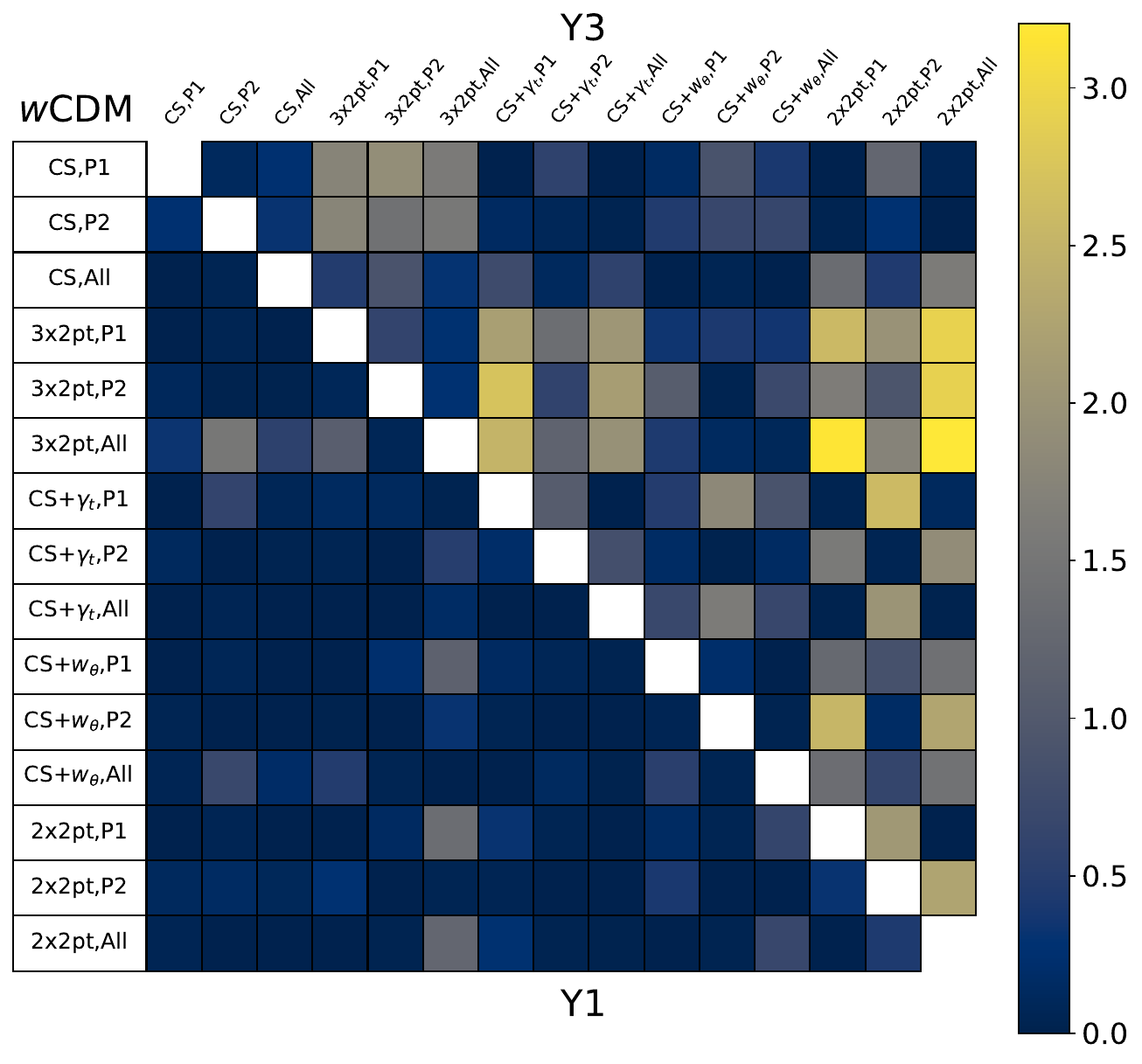}
\caption{Tensions between chains under split $\Lambda$CDM (left panel) and $w$CDM (right panel) models. As described in Sec.~\ref{sec:data_set_for_geometry}, the P1 external data combination is composed of Planck 2018 low-$\ell$ EE polarization data and the high-$\ell$ TTTEEE spectra truncated right after the first peak ($35<\ell<396$). On the other hand, P2 is the combination of SNIa + BAO + BBN. The 2x2pt + P2 chain is the only one with significant tension, against other DES-Y3 2PCFs, in $\Lambda$CDM split; the inflationary amplitude $A_\mathrm{s}$ seems to be the culprit of the observed tensions. When 2x2pt is combined with either P1 or All priors, we see lower tensions at the expense of degradation in goodness-of-fit (see Fig.~\ref{fig:210chi2plot}). The $w$CDM behavior is different; the 2x2pt + P1/\textsc{All} chains have the highest tensions, against 3x2pt+\textsc{All} caused by $\sigma_8^\mathrm{split}$, and there is no loss in goodness-of-fit compared with 2x2pt + P2 (see Fig.~\ref{fig:210chi2plot_wcdm}).}
\label{fig:internal_tension1} 
\end{figure*}

We evaluate tensions between different DES 2PCF combinations employing the parameter difference method on $\big\{ A_{\mathrm{s}}, n_{\mathrm{s}}, H_0,\Omega_\mathrm{m}^{\rm geo}, \sigma_8^{\rm split}(z=0) \big\}$ set of cosmological parameters, with an addition of $w^{\rm geo}$ in the split $w$CDM model. As a caveat, this metric does not model the existing correlations between the 2PCFs; the precise computation requires MCMC chains with repeated parameters, which is beyond our computational capabilities~\cite{raveri_non-gaussian_2021}. Figure~\ref{fig:internal_tension1} qualitatively indicates discrepancies; we see, for example, the well-known \textsc{redMaGiC} problems between 2x2pt and other probe combinations. Future utilization of machine learning emulators will allow the more precise calculation of tensions between the correlated DES 2PCFs with modest computational resources~\cite{Boruah:2022uac}. 

Interestingly, $A_\mathrm{s}$ appears to be the culprit of the observed tensions above two sigmas between 2x2pt and the remaining combinations of the DES-Y3 data vector. The highest observed tension in split $\Lambda$CDM happens between 2x2pt + P2 and $\xi_\pm + w_\theta$ + P1, entirely due to shifts on $A_\mathrm{s}$ as both chains favors $\Delta \Omega_{\rm m} < 0$. Figure~\ref{fig:210chi2plot} reveals that CMB priors degrade the goodness of fit to DES-Y3 2x2pt data by $\Delta \chi^2 \approx 5$. In all other DES-Y3 2PCFs, swapping P1 with P2 priors does not affect $\chi^2$ nearly as much. However, the detailed comparison between $\xi_\pm + w_\theta$ and $\xi_\pm + \gamma_t$ against 3x2pt stands out. Both $\xi_\pm + w_\theta$ and $\xi_\pm + \gamma_t$  combinations show virtually no $\chi^2$ changes between all three priors;the same is not true for 3x2pt as there is a $\Delta\chi^2 \approx 1.21$ degradation on DES goodness-of-fit when CMB data is present.

The behavior in $w$CDM split is different; growth parameters can restore the DES
2x2pt goodness-of-fit when $A_\mathrm{s}$ is set by the CMB prior, as shown in Fig.~\ref{fig:210chi2plot_wcdm}. The $A_\mathrm{s}$ tension between DES-Y3 2x2pt + P2 and DES-Y3 2x2pt + P1/All is also smaller on $w$CDM when compared with $\Lambda$CDM split. The DES-Y3 2x2pt predicts nonzero values for the principal component
$\Delta \Omega_\mathrm{m} - \Delta w$ for all external data combinations, the more extreme deviation from zero happening on DES-Y3 2x2pt + All chain. The better fit to DES 2x2pt makes such nonzero detection more meaningful than the $\Lambda$CDM split model. 

Finally, the left and right panels on Fig.~\ref{fig:internal_tension1} show that DES-Y3 2x2pt + P1/All chains have higher tension levels against other 2PCFs than DES-Y3 2x2pt + P2, the opposite of what we observe in $\Lambda$CDM split. Indeed, when cosmic shear is added to 2x2pt, the predicted $\sigma_8^\mathrm{split}$ value shifts by more than three sigmas. Unfortunately, both $\Lambda$ and $w$CDM split models have similar DES 3x2pt goodness-of-fit; growth parameters can't alleviate the incompatibility between galaxy-galaxy lensing and galaxy clustering in the 3x2pt chains (see Fig.~\ref{fig:210chi2plot} and~\ref{fig:210chi2plot_wcdm}).

\section{Conclusions}
\label{sec:Conclusion}

This paper studies  the growth-geometry split with DES-Y1 and DES-Y3 data in combination with external data sets. We utilize the \textsc{Cobaya-CosmoLike Architecture (Cocoa)} software to efficiently run a large number of MCMC chains that allow us to explore the variation of results for different probes and prior combinations. 

For DES-Y1 we find that $\Delta \Omega_{\rm m}$ in $\Lambda$CDM and $\Delta \Omega_{\rm m} - \Delta w$ in $w$CDM are both consistent with 0 for all permutations of DES 2PCFs and external prior combinations.

% We show DES-Y3 constraints on both $\Lambda$CDM and $w$CDM growth geometry splits. 
In the case of DES-Y3, we find that cosmic shear and 3x2pt results are consistent with equal geometry and growth parameters. Combining cosmic shear and galaxy-galaxy lensing also does not indicate deviations between growth and geometry parameters. However, both the $\xi_\pm + w_\theta$ and $\gamma_t + w_\theta$ combinations of 2PCF indicate $\Delta \Omega_{\rm m} < 0$ in $\Lambda$CDM and $\Delta \Omega_{\rm m} - \Delta w < 0$ in $w$CDM splits. These results hold with both P1 and P2 priors, which is interesting as they predict different values for the primordial power spectrum amplitude $A_\mathrm{s}$. In light of the well-known DES-Y3 problems of the \textsc{redMaGiC} sample, we do not interpret these results as a detection but rather assume that it is a residual of unsolved systematics. We plan to further explore this with alternative lens samples, in particular the \textsc{MagLim} sample, and when marginalizing over the $X_\mathrm{lens}$ ~\citep{2022PhRvD.106d3520P}.

Comparing our work with other results in the literature is unfortunately not straightforward since there are several different ways how $\Lambda$CDM parameters can be split into geometry and growth. This work focuses on additional parameters allowing an anomalous late-time growth-independent evolution of the matter power spectrum. In ~\cite{Andrade:2021njl}, on the other hand, the growth parameters also affect the source function of the CMB power spectrum. Thus, different values of $\Omega_\mathrm{m}^{\rm growth}$ affect both early and late-time dynamics and produce significant changes to the CMB temperature and polarization power spectra. These split parameterizations that affect both early and late-time dynamics produce $\Delta \Omega_{\rm m} \neq 0$ detections at a level greater than 4$\sigma$, much higher than what we observe with our adopted late-time scale-independent modifications to the matter power spectrum. ~\cite{Linder:2005in, Basilakos:2019hlb} describe a third possibility for the split. Their growth parameters affect the growth index $\gamma$, which is a single parameter that approximately describes the $\Lambda$CDM growth history in the late Universe.

Several extensions to this paper come to mind: Firstly, we already mentioned that it will be important to study the impact of other lens samples, in particular the \textsc{MagLim} sample. Secondly, additional cosmological information from external datasets, such as including more scales of the CMB temperature and polarization power spectrum, and adding CMB lensing are near-term extensions of this work. The $w$CDM split would also benefit from extra information on $w^\mathrm{geo}$ from the observed DES Type IA supernova included in the new \textsc{Phanteon+} sample~\cite{2022ApJ...938..113S}. Thirdly, we plan to include small-scale information to increase the constraining power on growth-geometry split parameters, e.g. by modeling baryons in cosmic shear as in \cite{hem20} or modeling galaxy bias in 2x2pt via effective field theory \cite{Kokron:2021xgh} or via Halo Occupation Distribution models \cite{Krause:2016jvl,2005ApJ...633..791Z,2011ApJ...736...59Z}. 

While this paper does not show any hints of new physics beyond $\Lambda$CDM, future datasets from Rubin Observatory's LSST~\cite{LSST:2008ijt}, the Roman Space Telescope~\cite{Dore:2019pld}, and the Euclid mission~\cite{laureijs2011euclid}, in combination with the Dark Energy Spectroscopic Instrument~\cite{2019BAAS...51g..57L}, Simons Observatory~\cite{2019JCAP...02..056A} and the CMB-S4 mission~\cite{2016arXiv161002743A} will significantly tighten the statistical error budget on cosmological models beyond $\Lambda$CDM and $w$CDM. It is now timely to develop the theoretical toolbox to efficiently and consistently explore these models across datasets.

\section*{Acknowledgements}

Simulations in this paper use High Performance Computing (HPC) resources supported by the University of Arizona TRIF, UITS, and RDI and maintained by the UA Research Technologies department. The authors would also like to thank Stony Brook Research Computing and Cyberinfrastructure, and the Institute for Advanced Computational Science at Stony Brook University for access to the high-performance SeaWulf computing system, which was made possible by a $\$1.4$M National Science Foundation grant ($\#1531492$). TE and JX are supported by the Department of Energy grant DE-SC0020215. EK is supported by the Department of Energy grant DESC0020247 and an Alfred P. Sloan Research Fellowship.

\bibliography{ref_short.bib}

\appendix

\end{document}